\documentclass[e-print,pra,twocolumn,amsmath,amssymb,superscriptaddress]{revtex4-1}
\usepackage{epsfig, graphicx,graphics,amsmath,amssymb,float}
\usepackage[T1]{fontenc}
\usepackage[latin9]{inputenc}
\usepackage{amsmath}
\usepackage{amssymb}
\usepackage{xcolor}
\usepackage{amscd}
\usepackage{bm}
\usepackage{bbold}
\usepackage{psfrag}

\usepackage{bbm} 

\usepackage[caption=false]{subfig}
\usepackage{subfig}
\usepackage{color}
\usepackage[bookmarks=true,colorlinks,linkcolor=blue,urlcolor=blue,citecolor=blue]{hyperref}

\newcommand{\be}{\begin{equation}}
\newcommand{\ee}{\end{equation}}
\newcommand{\bea}{\begin{eqnarray}}
\newcommand{\eea}{\end{eqnarray}}

\newcommand{\mc}{\mathcal}
\newcommand{\mb}{\mathbf}

\begin{document}

\title{Quadrupolar spin liquid, octupolar Kondo coupling and odd-frequency superconductivity  in an exactly solvable model}

\author{Carlene S. de Farias}

\affiliation{Instituto de F\'isica Gleb Wataghin, University of Campinas (Unicamp), Campinas, SP, 13083-859, 
Brazil}

\affiliation{International Institute of Physics and Departamento de F\'isica Te\'orica e Experimental, Universidade Federal do Rio Grande do Norte, Campus Universit\'ario, Lagoa Nova,  Natal, RN, 59078-970, Brazil}

\author{Vanuildo S. de Carvalho}
\affiliation{Instituto de F\'isica Gleb Wataghin, University of Campinas (Unicamp), Campinas, SP, 13083-859, 
Brazil}

\author{Eduardo Miranda}

\affiliation{Instituto de F\'isica Gleb Wataghin, University of Campinas (Unicamp), Campinas, SP, 13083-859, 
Brazil}

\author{Rodrigo G. Pereira}
\affiliation{International Institute of Physics and Departamento de F\'isica Te\'orica e Experimental, Universidade Federal do Rio Grande do Norte, Campus Universit\'ario, Lagoa Nova,  Natal, RN, 59078-970, Brazil}

\date{\today}
\begin{abstract}

We propose an exactly  solvable model for $j_{\text{eff}}=\frac32$ local moments on the honeycomb lattice.  Our construction is guided by a symmetry analysis and by the requirement of an exact solution in terms of a Majorana fermion representation for multipole operators. The main interaction in the model can be interpreted as a bond-dependent quadrupole-quadrupole interaction. When   time reversal symmetry  is explicitly broken, we obtain a gapped spin liquid    with a single chiral Majorana edge  mode. We also investigate another   solvable model in which  the time-reversal-invariant spin liquid is coupled to conduction electrons in a superconductor. In the presence of a Kondo-like  coupling that involves the octupole moment of the localized spins, the itinerant electrons hybridize with the emergent Majorana fermions in the spin liquid. This leads to spontaneous time reversal symmetry breaking and generates odd-frequency pairing. Our results suggest that   $j_{\text{eff}}=\frac32$ systems with strong quadrupole-quadrupole interactions may provide  a  route towards non-Abelian quantum spin liquids and unconventional superconductivity. 

\end{abstract}



\maketitle

\section{Introduction\label{sec:level1}}

Quantum spin liquid phases have fascinated condensed matter physicists since Anderson's proposal of resonating valence bond states  \cite{Anderson1973a,Anderson1987}. These phases   harbor exotic properties such as spin fractionalization  and long-range entanglement \cite{Savary_2016,Balents2010}. Unlike   classical magnetic phases  that   spontaneously break  symmetries of the Hamiltonian,  quantum spin liquids  are not characterized by  local order parameters; in fact,  their study  helped develop  the concept of  topological order \cite{Wen1995,Wen2002}. 

Kitaev's honeycomb model \cite{Kitaev2006} is the best known example of an exactly solvable model with  a quantum spin liquid ground state. The solution works by expressing the spin $S=\frac12$ operators in terms of  Majorana fermions and realizing that the exact  excitations correspond to deconfined Majorana  fermions    in the background of a static $\mathbbm{Z}_{2}$ gauge field. Experimentally, the bond-dependent anisotropic exchange interaction of the Kitaev model is realized  in the  iridates    (Na,Li)$_{2}$IrO$_{3}$ and   the  ruthenium compound $\alpha$-RuCl$_{3}$ \cite{Hermanns2018,Winter_2017,Takagi2019}, where it is  generated via the Jackeli-Khalliulin mechanism \cite{Khaliullin2005,Jackeli2009}. Essential  ingredients for the latter are the strong spin-orbit coupling of 4$d^5$ or 5$d^5$ magnetic ions and the environment of  edge-sharing octahedra formed by the ligands.  The Kitaev interaction, parametrized by coupling constant $K$, arises as the leading term in the effective spin model for  $j_{\textrm{eff}}=\frac12$  local moments. A more general   model  that takes into account subleading exchange paths and   trigonal distortions of the octahedra must also include the   Heisenberg interaction $J$ and  the  anisotropic interactions denoted $\Gamma$ and $\Gamma'$ \cite{Rau2014,Gordon2019}. As a matter of fact, the Kitaev model can  be regarded as an exactly solvable point in the parameter space of the  $J$-$K$-$\Gamma$-$\Gamma'$ model.   In the iridates  and   $\alpha$-RuCl$_{3}$,  the  additional   couplings  beyond the Kitaev model are large enough that these materials fall outside the Kitaev spin liquid phase and undergo magnetic ordering transitions at low temperatures \cite{Hermanns2018,Winter_2017,Takagi2019}.  

In the past few years, alternative routes to Kitaev magnetism have been explored. In particular, the search has been extended to systems with more degrees of freedom,   beyond the picture of $j_{\textrm{eff}}=\frac12$ moments. For instance, materials with $4d^1$ or $5d^1$ configuration are described by effective models with $j_{\textrm{eff}}=\frac32$ moments, which can be represented by  pseudospin and pseudo-orbital degrees of freedom  \cite{Chen2010,Natori2016,Natori2017,Romhanyi2017}. The higher value of  $j_{\textrm{eff}}$ allows for multipolar interactions  which can promote hidden multipolar orders or even quantum spin-orbital liquid phases \cite{Wang2009,WitczakKrempa2014,Natori2018,Yamada2018,Ishikawa2019}. Multipolar interactions also play an important role in rare-earth systems \cite{Baker1971,Shiina1997,Kubo2005,Santini2009,Lee2018}, where electrons in the $f$-shell are much more localized and   have stronger spin-orbit coupling than $d$ electrons in transition metal compounds. Indeed,  rare-earth magnets analogous  to   $j_{\textrm{eff}}=\frac12$ Kitaev materials  have   been proposed recently  \cite{Li2017,Jang2019,Xing2019}.   The spin-$S$ Kitaev model with $S>\frac12$, generated microscopically by Hund's coupling in the transition metal and strong spin-orbit coupling in the ligands,   has also attracted considerable attention   \cite{Koga2018,Oitmaa2018,Stavropoulos2019,Dong2019,Xu2020,Hickey2020}. Unlike the original  $S=\frac12$ Kitaev model \cite{Kitaev2006}, however, these   higher spin  models  are not integrable in general, and the characterization  of putative quantum spin liquid phases needs to rely on numerics or analytical mean-field approximations.  

In this paper, we propose  an exactly  solvable  $j_{\textrm{eff}}=\frac32$ model on the honeycomb lattice with a spin-orbital liquid ground  state and Majorana fermion excitations. The guiding principles behind our construction are the exact solvability, as in previous  generalizations of  the Kitaev model to higher-dimensional local Hilbert spaces  \cite{Yao2009,Wu2009,Yao2011,Dwivedi2018}, and a  symmetry-based analysis of the possible interactions between  $j_{\textrm{eff}}=\frac32$ moments in an octahedral  crystal  field. Whenever possible, we shall discuss the physical interpretation of the various terms in the integrable model, but  here we do not attempt to derive them from  detailed microscopic mechanisms for specific materials. However, our toy model can be regarded as an exactly solvable point in the parameter space of  more realistic  models for  $j_{\textrm{eff}}=\frac32$ systems. In this sense, the physical properties  discussed here may be relevant to $4d^1$ or $5d^1$ materials or rare-earth systems with a  $\Gamma_8$ quartet ground state \cite{Shiina1997,Abragam2012} if   realistic models for the latter happen to fall near the integrable point.  In fact, the leading term in our Hamiltonian  can be interpreted as a quadrupole-quadrupole interaction, which does appear as a symmetry-allowed  term in the model for $j_{\textrm{eff}}=\frac32$ honeycomb systems including the effects of Hund's coupling \cite{Natori2018}. We also consider a single-ion anisotropy  term that lowers the point group symmetry and  leads to a nonzero expectation value of the local quadrupole moment, but the spin-orbital excitations remain fractionalized into Majorana fermions. Upon breaking time reversal symmetry, we obtain fully  gapped spin-orbital excitations in the bulk, but   one gapless Majorana mode on the edge, characteristic of a non-Abelian phase with a quantized thermal Hall conductance \cite{Kitaev2006}.  

\begin{figure}[t]
\includegraphics[scale=0.44]{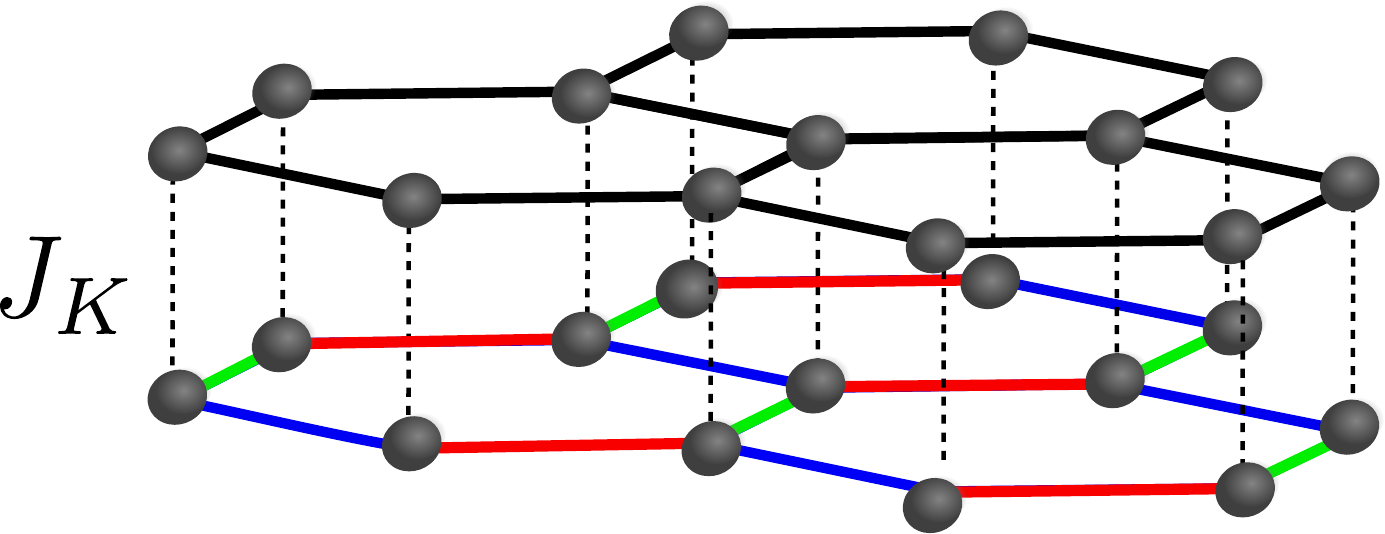}\caption{Schematic representation of the  honeycomb lattice containing
the  spin liquid (lower plane) and the itinerant  electron (upper plane) degrees of freedom. The $j_{\textrm{eff}}=\frac32$ local moments in the spin liquid interact  via  Kitaev-like quadrupole-quadrupole interactions $K_{\gamma}$, with $\gamma=x,y,z$ labeling the different nearest-neighbor bonds    indicated   by the colors red, blue and green, respectively. The conduction electrons are described by   nearest-neighbor   and   next-nearest-neighbor  couplings which include hopping  and pairing amplitudes. The two subsystems interact via  a  Kondo-like coupling $J_{K}$ that involves the octupole moments of the localized spins. \label{Honeycomb_Lattice} }
\end{figure}

Furthermore, we investigate the coupling of this ``quadrupolar spin liquid'' to itinerant electrons, as illustrated in Fig. \ref{Honeycomb_Lattice}.  This part is motivated by recent studies of the Kondo-Kitaev model within  mean-field approximations  \cite{Seifert2018,Choi2018} and by {\it ab initio} calculations of $\alpha$-RuCl$_3$/graphene heterostructures \cite{Biswas2019}, which  suggest that the Kondo coupling between  a Fermi liquid and a Kitaev spin liquid  can  give rise to exotic  superconductivity. Again we choose the interactions so that the total Hamiltonian is exactly solvable in terms of free Majorana fermions and conserved $\mathbbm Z_2$  bond  variables. This  requires the decoupled  electronic system to be a superconductor that breaks inversion and SU(2) spin rotational symmetry, but preserves time reversal symmetry. The  itinerant electrons and $j_{\textrm{eff}}=\frac32$ local moments interact via  an octupolar Kondo coupling, analogous to the coupling   studied in the context of heavy-fermion systems  where the magnetic ions form  non-Kramers doublets \cite{Zhang2018,Patri2019}.  We find that the hybridization between Majorana fermions in the spin liquid and in the superconductor leads to  \emph{spontaneous} breaking of time reversal symmetry. Within the exactly solvable model, the ground state is   degenerate between the choices of uniform (``ferro'') or staggered (``antiferro'') hybridization order parameters. In both cases, the effective action for the electrons in the superconductor contains odd-frequency pairing \cite{Berezinskii1974,Kirkpatrick1991,Balatsky1992,Coleman1993,Coleman1994,Belitz1999,Balatsky2019,Tanaka2012}. We then analyze the effects of weak integrability-breaking perturbations that lift this degeneracy, and obtain either a  gapped    superconductor with chiral edge states or a gapless superconductor with a Bogoliubov Fermi surface \cite{Brydon2018} and antichiral edge states \cite{Franz2018}.

The paper is organized as follows. First, in Sec. \ref{sec:The-model}, we introduce the exactly solvable model with quadrupole-quadrupole interactions between local moments.  In Sec. \ref{sec:kondo}, we consider the full model with the  Kondo-like coupling  between the spin  liquid and the superconductor. We discuss   properties of the spectrum and the spontaneous time reversal symmetry breaking in the coupled system. In Sec. \ref{sec:Instabilities-induced-by}, we calculate the effective action for the electrons in the superconductor after exactly integrating out the Majorana  fermions associated with the local moments. In Sec. \ref{subsec:Perturbations}, we analyze the effects of  perturbations to  the exactly solvable model.  Finally, we present our conclusions   in Sec. \ref{sec:Discussions-and-conclusions}.

\section{Exactly solvable quadrupolar spin liquid model \label{sec:The-model}}

\subsection{Symmetry considerations}

Consider local moments with effective total angular momentum $j_{\textrm{eff}}=\frac32$. For concreteness, we can think of transition metal  ions with a $4d^1$ or $5d^1$ configuration subject to  a strong  crystal field of ligand octahedra and to strong  spin-orbit coupling. In this case, the single electron in the open $d$ shell occupies  the  threefold degenerate $t_{2g}$ orbitals with effective orbital angular momentum $l_{\textrm{eff}}=1$. The spin-orbit coupling splits the energy levels into a higher-energy $j_{\textrm{eff}}=\frac12$ doublet and the low-lying $j_{\textrm{eff}}=\frac32$ quadruplet   \cite{Winter_2017}. Alternatively, we may consider rare earth ions   which   have a quartet ground state in an octahedral crystal field \cite{Shiina1997,Kubo2005}.

\begin{table*}
\caption{\label{tab:multipoles}Local operators   acting on  $j_{\textrm{eff}}=\frac32$ states. Overlines indicate the
symmetrization with respect to   permutations of the
indices, e.g., $\overline{J^{x}(J^{y})^{2}}=J^{x}(J^{y})^{2}+J^{y}J^{x}J^{y}+(J^{y})^{2}J^{x}$. Here $\mathbf u^\gamma$ and $\mathbf v^\gamma$ are unit vectors in the $xz$ plane given by $\mathbf u^x=-\frac12\hat{\mathbf z}+\frac{\sqrt3}2\hat{\mathbf x}$, $\mathbf u^y=-\frac12\hat{\mathbf z}-\frac{\sqrt3}2\hat{\mathbf x}$, $\mathbf u^z=\hat{\mathbf z} $,  $\mathbf v^x=-\frac{\sqrt3}2\hat{\mathbf z}-\frac{1}2\hat{\mathbf x}$, $\mathbf v^y=\frac{\sqrt3}2\hat{\mathbf z}-\frac{1}2\hat{\mathbf x}$, $\mathbf v^z=\hat{\mathbf x} $. 
Adapted from Refs. \cite{Shiina1997,Natori2017}.}

\begin{tabular}{ccccc}
\hline 
\hline 

Moment  & Symmetry  & Operators  & \qquad$\textbf{s},\boldsymbol{\tau}$ representation \qquad\qquad  &Majorana    representation \tabularnewline
\hline 
Dipole  & $\Gamma_{4}$  & $J^{x}$  & $-s^{x}(1+4\mathbf u^x\cdot \boldsymbol \tau)$ &$\frac{i}2\eta^y\eta^z-2i\eta^x\left(-\frac12\theta^z+\frac{\sqrt3}2\theta^x\right)$\tabularnewline
 &  & $J^{y}$  & $-s^{y}(1+4\mathbf u^y\cdot \boldsymbol \tau)$ &$\frac{i}2\eta^z\eta^x-2i\eta^y\left(-\frac12\theta^z-\frac{\sqrt3}2\theta^x\right)$\tabularnewline
 &  & $J^{z}$  & $-s^{z}(1+4\mathbf u^z\cdot \boldsymbol \tau)$  
 &$\frac{i}2\eta^x\eta^y-2i\eta^z\theta^z$\tabularnewline
\hline 
Quadrupole  & $\Gamma_{3}$  & $O^{3z^2-r^2}=\frac1{3}[3(J^{z})^{2}-\textbf{J}^{2}]
$  & $2\tau^{z}$ &$-i\theta^x\theta^y$
\tabularnewline
 &  & $O^{x^2-y^2}=\frac1{\sqrt3}[(J^{x})^{2}-(J^{y})^{2}]$  & $2\tau^{x}$  
 &$-i\theta^y\theta^z$\tabularnewline 
 & $\Gamma_{5}$  & $O^{xy}=\frac1{\sqrt3}\overline{J^{x}J^{y}} $  & $-4s^{z}\tau^{y}$ & $i\eta^z\theta^y$\tabularnewline
 &  & $O^{yz}=\frac1{\sqrt3}\overline{J^{y}J^{z}} $  & $-4s^{x}\tau^{y}$&$i\eta^x\theta^y$ \tabularnewline
 &  & $O^{zx}=\frac1{\sqrt3}\overline{J^{x}J^{z}} $  & $-4s^{y}\tau^{y}$ & $i\eta^y\theta^y$\tabularnewline
\hline 
Octupole  & $\Gamma_{2}$  & $T^{xyz}= \frac{2}{3\sqrt{3}}\overline{J^{x}J^{y}J^{z}}$  & $2\tau^{y}$ &$-i\theta^z\theta^x$\tabularnewline
 & $\Gamma_{4}$  & $\frac23 (J^{x})^{3}-\frac{1}{3}(\overline{J^{x}(J^{y})^{2}}+\overline{(J^{z})^{2}J^{x}})$  & $-2s^{x}(1-\mathbf u^x\cdot \boldsymbol \tau)$&$i\eta^y\eta^z+\frac{i}2\eta^x\left(-\frac12\theta^z+\frac{\sqrt3}2\theta^x\right)$ \tabularnewline
 &  & $ \frac23(J^{y})^{3}-\frac{1}{3}(\overline{J^{y}(J^{z})^{2}}+\overline{(J^{x})^{2}J^{y}})$  & $-2s^{y}(1-\mathbf u^y\cdot \boldsymbol \tau)$&$i\eta^z\eta^x+\frac{i}2\eta^y\left(-\frac12\theta^z-\frac{\sqrt3}2\theta^x\right)$ \tabularnewline
 &  & $\frac23(J^{z})^{3}-\frac{1}{3}(\overline{J^{z}(J^{x})^{2}}+\overline{(J^{y})^{2}J^{z}})$  & $-2s^{z}(1-\mathbf u^z\cdot \boldsymbol \tau)$ &$i\eta^x\eta^y+\frac{i}2\eta^z\theta^z$\tabularnewline
  
 & $\Gamma_{5}$  & $ \frac{2}{3\sqrt3}[\overline{J^{x}(J^{y})^{2}}-\overline{(J^{z})^{2}J^{x}}]$  & $-4s^{x}\mathbf v^x\cdot \boldsymbol \tau$ & $-i\eta^x\left(-\frac12\theta^x-\frac{\sqrt3}2\theta^z\right)$ \tabularnewline
 &  & $ \frac{2}{3\sqrt3}[\overline{J^{y}(J^{z})^{2}}-\overline{(J^{x})^{2}J^{y}}]$  & $-4s^{y}\mathbf v^y\cdot \boldsymbol \tau$ & $-i\eta^y\left(-\frac12\theta^x+\frac{\sqrt3}2\theta^z\right)$\tabularnewline
 &  & $ \frac{2}{3\sqrt3}[\overline{J^{z}(J^{x})^{2}}-\overline{(J^{y})^{2}J^{z}}]$  & $-4s^{z}\mathbf v^z\cdot \boldsymbol \tau$& $-i\eta^z\theta^x$ \tabularnewline
  \hline
  \hline
\end{tabular}
\end{table*}

The local Hilbert space is spanned by the four eigenstates  of the $J^z$   operator, $J^z|m_J\rangle=m_J|m_J\rangle$, with $m_J=\pm \frac12,\pm\frac32$.    These states transform under rotations as the $\Gamma_8$ representation of the octahedral double  group \cite{Abragam2012}. The  operators acting  in the local Hilbert space can be organized  into dipole, quadrupole and octupole moments \cite{Shiina1997,Santini2009} as shown in Table \ref{tab:multipoles}.  While the dipole and octupole moments involve odd powers of  components of $\mathbf J$ and change sign under time reversal, the quadrupole moments  contain even powers and are time-reversal invariant.  

We rewrite the eigenstates of $J^z$ in terms of two pseudospin-1/2 quantum numbers in the form $|s^z,\tau^z\rangle$ \cite{Shiina1997,Natori2016}, identifying   
\bea
&\left|m_J= \frac32\right\rangle=\left|s^z=-\frac12,\tau^z=\frac12\right\rangle\equiv \left|-+\right\rangle,\nonumber\\
&\left|m_J=\frac12\right\rangle=-\left|s^z=\frac12,\tau^z=-\frac12\right\rangle\equiv-\left|+-\right\rangle,\nonumber\\
&\left|m_J=-\frac12\right\rangle=\left|s^z=-\frac12,\tau^z=-\frac12\right\rangle\equiv\left|--\right\rangle,\nonumber\\
&\left|m_J= -\frac32\right\rangle=-\left|s^z=\frac12,\tau^z=\frac12\right\rangle\equiv-\left|++\right\rangle. 
\eea
We refer to $s^z$ and $\tau^z$ as the pseudospin and pseudo-orbital quantum numbers, respectively. The operators $\mathbf s$ and $\boldsymbol\tau$ acting on the corresponding degrees of freedom obey the    algebra $[s^\alpha,s^\beta]=i\epsilon^{\alpha\beta\gamma}s^\gamma$, $[\tau^\alpha,\tau^\beta]=i\epsilon^{\alpha\beta\gamma}\tau^\gamma$, and $[s^\alpha,\tau^\beta]=0$. In this notation, the time reversal operator is written as $T=-2is^y\mc K$, where $\mc K$ denotes complex conjugation.  Thus, two states with the same $\tau^z$ eigenvalue  form a  Kramers pair. States in a non-Kramers pair, having the same $s^z$ but      different $\tau^z$,      are associated with different electronic density profiles \cite{Natori2017}. The pseudospin and pseudo-orbital operators transform under time reversal as follows:\bea
T(s^x,s^y,s^z)T^{-1}&=&(-s^x,-s^y,-s^z),\nonumber\\
T(\tau^x,\tau^y,\tau^z)T^{-1}&=&(\tau^x,-\tau^y,\tau^z).
\eea
The representation of the multipole operators  in terms of $\mathbf s$ and $\boldsymbol\tau$ is shown in Table \ref{tab:multipoles}.   We note in particular the  transformation of $\mb s$ and $\boldsymbol\tau$ under a C$_3$  rotation around the [111] axis:
\bea
C_3(s^x,s^y,s^z)C_3^{-1}&=&(s^y,s^z,s^x),\nonumber\\
C_3(\tau^x,\tau^z)C_3^{-1}&=&\left(\frac{-\tau^x-\sqrt3\tau^z}2,\frac{-\tau^z+\sqrt3\tau^x}2\right),\nonumber\\ 
C_3 \tau^yC_3^{-1}&=&\tau^y.\label{C3rotation}
\eea
Importantly, the $\tau^y$ operator is invariant  under all rotations of the octahedral group, but changes sign under time reversal.  It corresponds to   the octupole moment  $T^{xyz}\propto\overline{J^xJ^yJ^z}$, where the overline indicates a sum over permutations of the indices, see Table \ref{tab:multipoles}.  By contrast, $\tau^x$ and $\tau^z$ are associated with   quadrupole moments, and C$_3$ rotations act as $120^\circ$ rotations of   $(\tau^z,\tau^x)$, in analogy with quantum compass models \cite{Nussinov2015}. 

\subsection{Time-reversal-invariant spin model\label{sec:THs}}

We begin with the model
\bea
H_s=\sum_{\gamma=x,y,z}\sum_{\langle ij\rangle_\gamma}K_\gamma O^{\alpha\beta}_iO^{\alpha\beta}_j- \lambda\sum_j O_j^{3z^2-r^2},\label{Hspin}
\eea
where  $\langle ij\rangle_\gamma$ labels a nearest-neighbor bond along the $\gamma $ direction, see Fig. \ref{Honeycomb_Lattice}, and $O_j^{3z^2-r^2}$ and  $O_j^{\alpha\beta}$ [with $(\alpha,\beta,\gamma)$ a cyclic permutation of $(x,y,z)$]    are quadrupole operators defined in Table \ref{tab:multipoles}. The first term amounts to a quadrupole-quadrupole interaction often invoked  in models for $f$-electron systems, where it stems from  electrostatic or phonon-mediated interactions \cite{McMahon1964,Baker1971,Shiina1997,Abragam2012}.   In $4d^1/5d^1$ systems, this interaction can also be generated  by   the coupling $(L_{i}^{\alpha}L_{i}^{\beta}+L_{i}^{\beta}L_{i}^{\alpha})(L_{j}^{\alpha}L_{j}^{\beta}+L_{j}^{\beta}L_{j}^{\alpha})$ 
\cite{Khaliullin2002}, where $\mathbf{L}_{i}$ is the $l_{\textrm{eff}}=1$ orbital angular momentum operator of the $t_{_{2g}}$ states,  upon projection onto the $j_{\textrm{eff}}=\frac32$ multiplet. In fact, the isotropic quadrupole-quadrupole interaction, with $K_x=K_y=K_z$,  appears in the effective spin model for $j_{\textrm{eff}}=\frac32$ systems on tricoordinated lattices with edge-sharing octahedra  \cite{Natori2018}.  The $\lambda$ term in Eq. (\ref{Hspin}) breaks the 
C$_3$ rotational symmetry even in the isotropic  case. This     single-ion anisotropy term can be associated with a distortion of the local  octahedral environment, which  lifts the  degeneracy between non-Kramers pairs  \cite{Chen2010}. 

To see why model (\ref{Hspin}) is exactly solvable, we introduce a Majorana fermion representation for spin  3/2  \cite{Wang2009,Yao2009}. In terms of pseudospin and pseudo-orbital operators, we write  \bea
s_j^\gamma&=&-\frac{i}4\epsilon^{\alpha\beta\gamma}\eta_j^\alpha\eta_j^\beta,\nonumber\\
\tau_j^\gamma&=&-\frac{i}4\epsilon^{\alpha\beta\gamma}\theta_j^\alpha\theta_j^\beta,
\eea
where  $\eta_j^\alpha$ and $\theta_j^\alpha$ are Majorana fermion operators that satisfy the anticommutation relations $\{\eta_j^{\alpha},\eta_l^\beta\}=\{\theta_j^{\alpha},\theta_l^\beta\}=2\delta_{jl}\delta^{\alpha\beta}$ and $\{\eta_j^{\alpha},\theta_l^\beta\}=0$. Similarly to  the original spin-1/2 Kitaev model \cite{Kitaev2006}, this Majorana fermion representation brings about a $\mathbbm Z_2$ gauge structure, since  physical observables are invariant under  $(\eta_j^\alpha,\theta_j^\alpha)\mapsto (-\eta_j^\alpha,-\theta_j^\alpha)$ . This gauge redundancy  enlarges the Hilbert space. To restrict states to the physical Hilbert space, we must impose the local constraint\be
D_j \equiv\; i\eta_j^x\eta_j^y\eta_j^z\theta_j^x\theta_j^y\theta_j^z=1\qquad \forall j. \label{constraint}
\ee
This can be implemented to all lattice sites by the projection operator $\mathcal{P} \equiv \prod_j [(D_j + 1)/2]$. The physical states $|\psi_\text{phys}\rangle$ of the quadrupolar spin liquid is obtained as $|\psi_\text{phys}\rangle = \mathcal{P}|\psi_0\rangle$, where $|\psi_0\rangle$ denotes the wave function of the Majorana fermions. The multipole operators are rewritten in terms of Majorana fermions as given in Table \ref{tab:multipoles}. In this representation, time reversal symmetry is implemented as complex conjugation, $TiT^{-1}=-i$, combined with  \bea
T\theta_j^{y}T^{-1}=-\theta_j^y,
\eea
leaving the other Majorana fermions invariant;  see Appendix \ref{app:Majorana}. The C$_3$ rotation acts as a cyclic permutation of $(\eta^x,\eta^y,\eta^z)$ and as a 120$^\circ$ rotation   $(\theta^z,\theta^x)\mapsto (-\frac12\theta^z+\frac{\sqrt3}{2}\theta^x,-\frac12\theta^x-\frac{\sqrt3}{2}\theta^z)$, analogous to Eq. (\ref{C3rotation}). Essentially, the ``scalar'' Majorana fermion $\theta^y$ inherits the symmetry properties of  the $\tau^y$ operator.

We rewrite the Hamiltonian Eq. (\ref{Hspin}) in terms of Majorana fermions according to Table \ref{tab:multipoles} and obtain\begin{equation}
H_{s}=i\sum_{\gamma}\sum_{\langle jl\rangle_{\gamma}}K_\gamma\hat{u}_{\langle jl\rangle_\gamma}\theta^y_{j}\theta_{l}^{y}+i\lambda\sum_{j}\theta_{j}^{x}\theta_{j}^{y},
\label{H_spins_majorana}
\end{equation}
where $\hat{u}_{\langle jl\rangle_\gamma}=-i\eta_{j}^{\gamma}\eta_{l}^{\gamma}$ are antisymmetric $\mathbbm Z_2$ bond operators, obeying $(\hat{u}_{\langle jl\rangle_\gamma})^2=1$  and $\hat{u}_{\langle jl\rangle_\gamma}=-\hat{u}_{\langle lj\rangle_\gamma}$, which commute with one another and with $H_s$. Once we fix the values of the conserved $\hat{u}_{\langle jl\rangle_\gamma}=\pm1$, the Hamiltonian becomes quadratic in the remaining Majorana fermions. The ground state is in the sector with zero $\mathbbm Z_2$ flux \cite{Lieb1994}, defined as the product of $\hat{u}_{\langle jl\rangle_\gamma}$ around each hexagon. We then set $\hat{u}_{\langle jl\rangle_\gamma}=1$ for all sites $j$ in sublattice A and $l$ the corresponding nearest neighbors in sublattice B. We have verified using exact diagonalization for small cluster systems that this condition gives the exact ground state energy of $H_s$. The resulting Hamiltonian for $\theta^x$ and $\theta^y$ is translationally invariant and can be diagonalized by a Fourier transform. Here we use the notation $\theta_j^\gamma \equiv \theta^\gamma_b(\mb R)$, where $\mb R$ is the position of the unit cell and  $b\in\{\textrm{A},\textrm{B}\}$ is a sublattice index. We can then write \be
\theta_j^\gamma=\sqrt{\frac2N}\sum_{\mathbf k\in\frac12\textrm{BZ}} \left[e^{i\mathbf k\cdot \mathbf R} \theta^\gamma_{b}(\mathbf k)+e^{-i\mathbf k\cdot \mathbf R} \theta^\gamma_{b}(-\mathbf k)\right],
\ee
where   the fermion operators in momentum space obey $\theta^\gamma_{b}(-\mathbf k)=[\theta^\gamma_{b}(\mathbf k)]^\dagger$, and the summation runs over half of the Brillouin zone. We find the dispersion relations  \begin{equation}
E_{s}(\mathbf{k})=\pm\frac{1}{\sqrt{2}}\sqrt{|g(\mathbf{k})|^{2}+2\lambda^{2}\pm|g(\mathbf{k})|\sqrt{|g(\mathbf{k})|^{2}+4\lambda^{2}}},  \label{dispHs}
\end{equation}
where $g(\mathbf{k})=K_{x}e^{i\mathbf{k}\cdot\mathbf{n}_{1}}+K_{y}e^{i\mathbf{k}\cdot\mathbf{n}_{2}}+K_{z}$.  
Here, $\mathbf{n}_{1}=\frac{\sqrt{3}a}{2}(1,\sqrt{3})$ and  $\mathbf{n}_{2}=\frac{\sqrt{3}a}{2}(-1,\sqrt{3})$ are the primitive lattice vectors in the $xy$ plane and we set $a=1$

\begin{figure}
\centering
\includegraphics[scale=0.35]{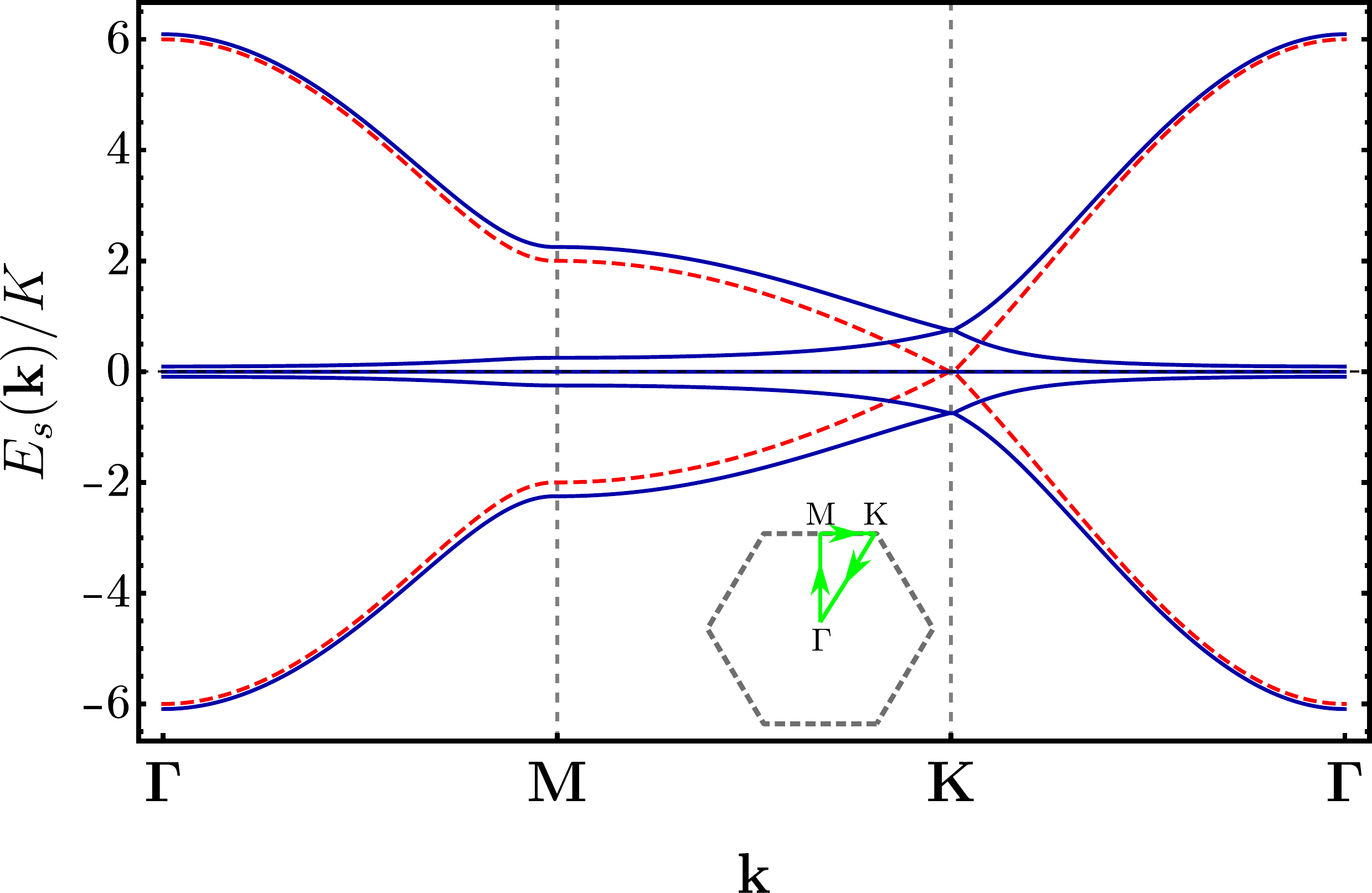}
\caption{Dispersion relation of the $\theta$ Majorana fermions in the quadrupolar spin liquid  along the high symmetry directions on the first Brillouin zone. Here we consider isotropic interactions, $K_{\gamma}=K$, and  two values of  the single-ion anisotropy parameter: $\lambda=0$ (dashed line) and $\lambda/K=0.75$ (solid line).
\label{Bulk_Dispersions_SL} }
\end{figure}
 
Having been absorbed into the bond operators, the $\eta^\gamma$ Majorana fermions are gapped out since it costs a finite energy to create $\mathbbm Z_2$ vortices \cite{Kitaev2006}. 
Figure \ref{Bulk_Dispersions_SL} shows  the dispersion relations  for $\theta$ fermions given by Eq. (\ref{dispHs}). Hereafter we focus on isotropic interactions,  $K_\gamma=K$. Note that for $\lambda=0$ we have two flavors of Majorana fermions, $\theta^x$ and $\theta^z$, which do not appear in the Hamiltonian. In terms of  $\boldsymbol \tau$ operators, we have a local U(1) symmetry, $[\tau^y_j,H_s]=0$ for $\lambda=0$. In this case, the $\theta^y$ mode displays  a gapless Dirac dispersion at the K point of the Brillouin zone, whereas $\theta^x$ and $\theta^z$  modes give rise  to zero energy flat bands, see Fig.  (\ref{Bulk_Dispersions_SL}). At each site,    $\theta^x_j$ and $\theta^z_j$  can be combined into a single complex fermion which  commutes with $H_s$, which implies that the ground state degeneracy increases exponentially with system size.

For $\lambda\neq0$, an energy gap opens up for the modes   coming  from the hybridization of $\theta^x$ and $\theta^y$, see Fig. \ref{Bulk_Dispersions_SL}.  The anisotropy associated with $\lambda$   lowers the ground state energy  and gaps  out  the excitations created by the pseudo-orbital operator $\boldsymbol\tau$. However,   the $\theta^z$  Majorana fermions still commutes with $H_s$. 
As a consequence,  the ground state of the quadrupolar spin liquid is   highly degenerate even  for $\lambda\neq 0$. The relation between the local conservation laws and the ground state degeneracy is discussed in more detail in App. \ref{App:degeneracy}. In Sec. \ref{sec:kondo} we will couple the Hamiltonian in Eq. (\ref{Hspin}) to conduction electrons in a superconductor. We shall see  that this coupling generates a dispersion for all modes and removes the exponential degeneracy in the ground state of the total system.  Alternatively, we can obtain an exactly solvable spin model with a unique ground state by  breaking time reversal symmetry, as we shall discuss in the next subsection.

\subsection{Breaking time reversal symmetry}

In  analogy with the Kitaev model in the presence of a magnetic field \cite{Kitaev2006}, we now investigate the effects of time reversal symmetry breaking in the quadrupolar spin liquid. We break time reversal while preserving the integrability of the model by adding to the Hamiltonian in Eq. (\ref{Hspin}) the following interactions:\be
\delta H_s=\kappa \sum_{\langle ij\rangle_\alpha\langle jk\rangle_\beta}O_i^{\beta\gamma}s^\gamma_jO_k^{\alpha\gamma}-\kappa'\sum_j T_j^{xyz}.
\ee
We propose the two terms in $\delta H_s$ based only on symmetry and exact solvabilty, but it is worth noting that they could in principle arise  as effective interactions generated by perturbation theory in a combination of a magnetic field, which couples to the time-reversal-odd pseudospin $s^\gamma$ contained in the dipole operator $J^\gamma$,  and strain fields which couple to the  quadrupole operators $O^{\alpha\beta}$. The small parameter in this case is  the ratio of the fields to  the  energy gap for changing the flux configuration by creating $\mathbbm Z_2$ vortices.  Using the Majorana fermion representation, we obtain\be
\delta H_s=i\kappa \sum_{\langle ij\rangle_\alpha\langle jk\rangle_\beta}\hat{u}_{\langle ij\rangle_\alpha}\hat{u}_{\langle jk\rangle_\beta}\theta_i^y\theta_k^y+i\kappa'\sum_{j}\theta_j^z\theta_j^x. 
\ee
We can then fix the values of the bond variables as discussed in Sec. \ref{sec:THs} and obtain a quadratic Hamiltonian $H_s+\delta H_s$ which involves all three $\theta^\gamma$ Majorana fermions.  We assume here that both $\kappa$ and $\kappa'$ are much smaller than the other energy scales of the model, in order to not change the ground-state   configuration of the bond variables.

\begin{figure}
\centering
\includegraphics[width=0.9\columnwidth]{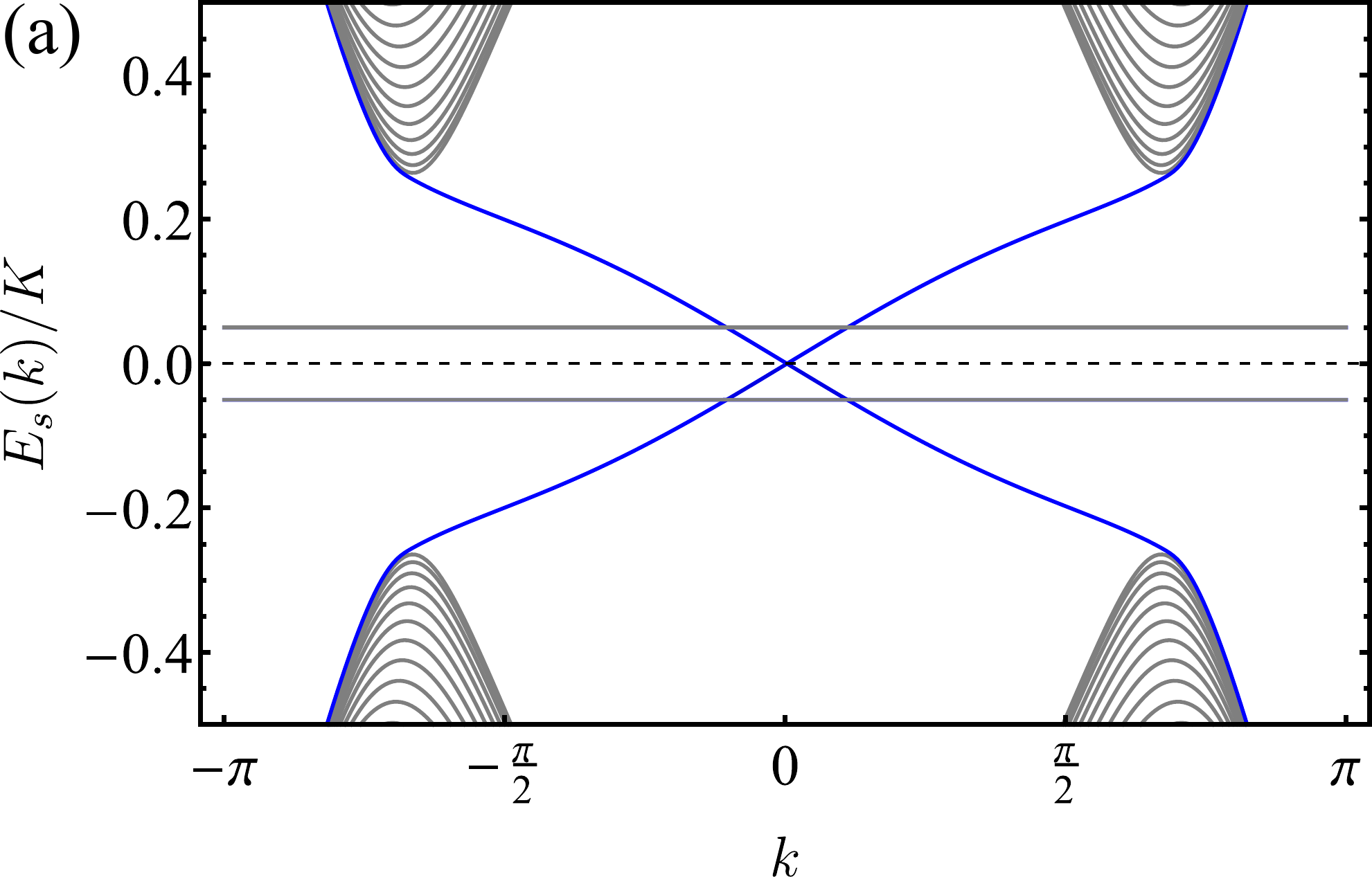}\vspace{.5cm}
\includegraphics[width=0.9\columnwidth]{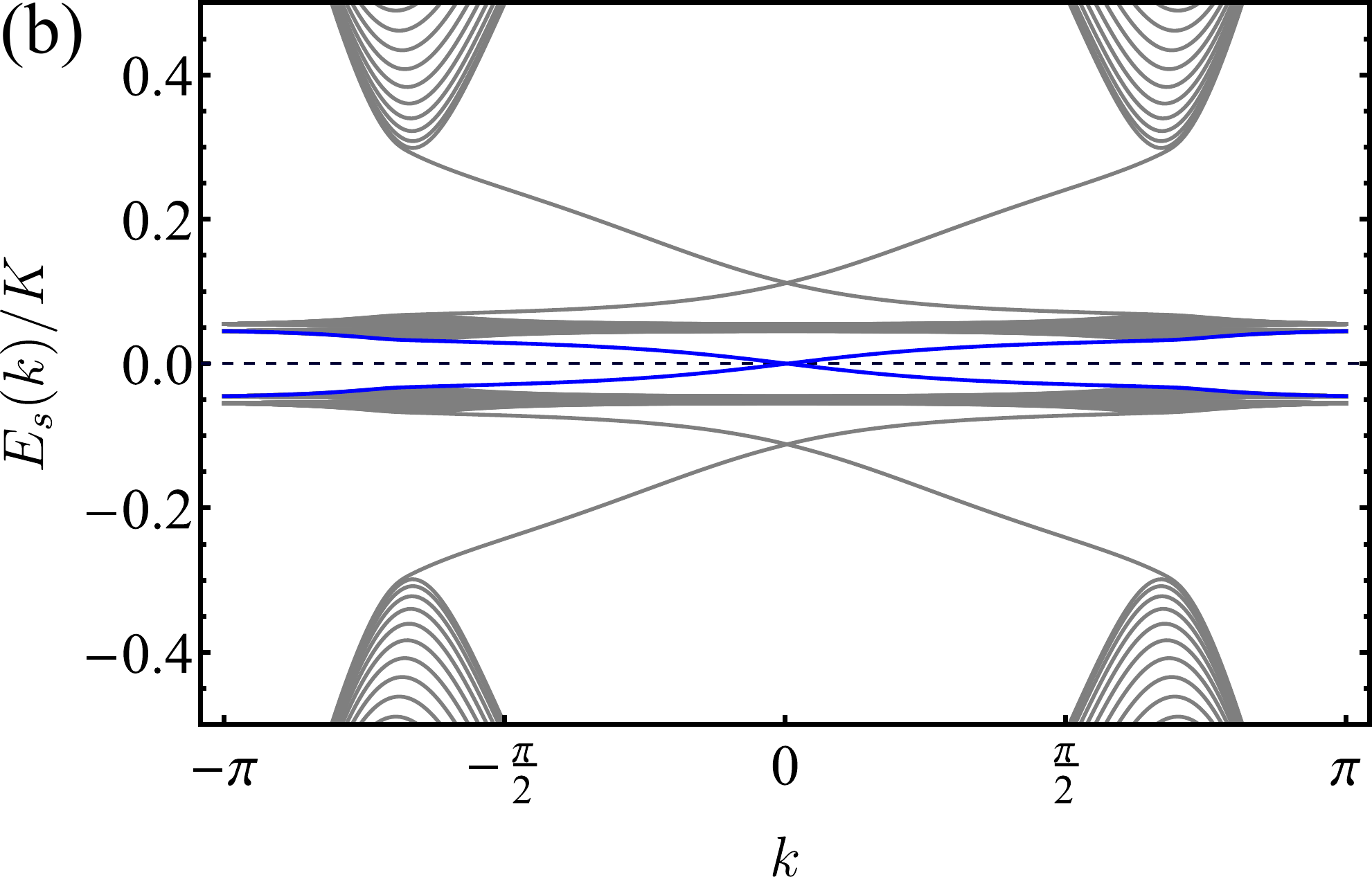}
\caption{Band structure of  the gapped spin liquid with broken time reversal symmetry on a strip with zigzag edges  and width  $W=80$ unit cells. Here we set $\kappa=\kappa'=0.05  K$ and consider two values of $\lambda$: (a) $\lambda=0$; (b) $\lambda=0.1K$. In both cases, the dispersion of the edge mode crosses zero  energy at a single point in the projection of the  Brillouin zone. 
\label{edgechiralSL} }
\end{figure}

In particular, for the  isotropic case   $\lambda=0$, the Majorana fermions $\theta^x$ and $\theta^z$ are decoupled from $\theta^y$. The ground state in this case  is an eigenstate of the local pseudo-orbital operators $\tau^y_j$ with eigenvalue  $\tau_j^y=\frac12$  for $\kappa'>0$. The $\kappa$ term gaps out the dispersion  of the $\theta^y$ Majorana fermions. Similarly to the Kitaev model   \cite{Kitaev2006}, we expect this gapped spin liquid to be a topological phase.  In fact, we find that the Chern number in this situation evaluates to $C = \pm 1$. We also obtain that, out of the three occupied bands of the $\theta^y$ Majorana fermions, only the one in the bottom of the spectrum contributes to $C$. To check this, we  compute the energy spectrum of the model on a strip geometry with open boundary conditions in the $y$ direction. In Fig. \ref{edgechiralSL}, we see that the spectrum contains  one  chiral Majorana  edge mode, equivalent to a topological superconductor with Chern number  $C=\pm1$. This result  is consistent with a recent conjecture for  spin-$S$ Kitaev spin liquids \cite{Hickey2020}, which states that when time-reversal symmetry is broken Kitaev spin liquids with half-integer spin $S$ exhibit a non-Abelian phase, while those with integer   $S$ possess, in contrast, an Abelian one. Our $j_{\text{eff}}=\frac32$ model is analogous to the $S=\frac32$ Kitaev model in the sense of a four-dimensional local Hilbert space, and indeed we find that  the Chern number is odd, corresponding to a non-Abelian phase with quantized thermal Hall conductance $\kappa_{xy}/T=\pi/12$ \cite{Kitaev2006}. 

We also note  that for $\lambda\neq0$ the eigenstates of the Hamiltonian are no longer eigenstates of $\tau_j^y$. In this case, the coupling of $\theta^{x,z}$ to $\theta^y$ turns the dispersionless modes seen in Fig. \ref{edgechiralSL}(a) into a band of mobile pseudo-orbital excitations as shown in Fig. \ref{edgechiralSL}(b).  As  expected, we also find $C = \pm 1$ for the Chern number. In contrast with the $\lambda = 0$ behavior, all occupied bands acquire here a non-trivial band topology, contributing with $\pm 1$ to this value of the Chern number.

\section{Coupling the spin liquid to a superconductor \label{sec:kondo}}

We now turn to  the Hamiltonian
\begin{equation}
H=H_{s}+H_{c}+H_{K}.\label{eq:KitaevKondo}
\end{equation}
Here, $H_s$  describes the quadrupolar spin liquid    in Eq. (\ref{Hspin}), and we shall assume $\lambda\neq0$. The new terms, $H_c$ and $H_K$, describe, respectively,  the conduction electrons in the superconductor and the   Kondo coupling between electrons and   $j_{\textrm{eff}}=\frac32$ local moments.  Our goal in this section is to write down $H_c$ and $H_K$ that produce  an exactly solvable model without any zero-energy flat bands. We shall see that the Kondo coupling gives rise to time-reversal-symmetry-breaking superconductivity.

\subsection{Time-reversal-invariant superconductor\label{Time-reversal_symmetric_SC}}
 
\begin{figure}
\centering
\includegraphics[width=0.55\columnwidth]{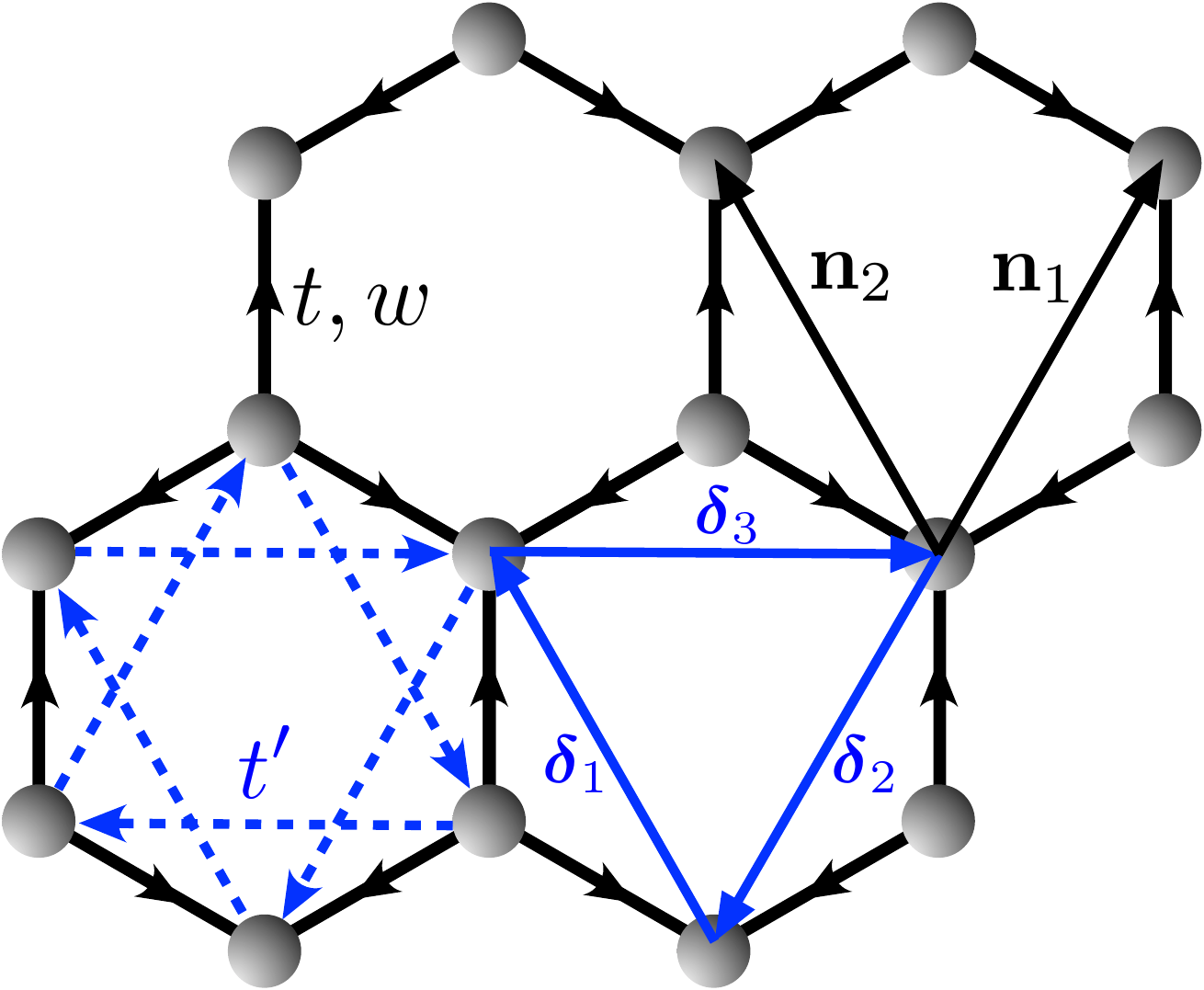}
\caption{Couplings in the Hamiltonian for itinerant electrons. Nearest-neighbor bonds with couplings $t$ and $w$ are oriented from the A to the B sublattice. Next-nearest-neighbor bonds with coupling $t'$ are oriented as indicated by the blue dashed arrows. 
\label{Haldane} }
\end{figure}

The following Bogoliubov-de-Gennes Hamiltonian describes the itinerant electron system:
\bea
H_{c}&=&i\sum_{\langle jl\rangle}\Psi_{j}^{\dagger}\left[t_{jl}(\sigma^x\rho^z+\sigma^z\rho^x)+w_{jl}(\sigma^z\rho^z-\sigma^x\rho^x)\right]\Psi_{l}^{\phantom\dagger}\nonumber\\
&&+i\sum_{\langle\langle jl\rangle\rangle} t'_{jl}\Psi_{j}^{\dagger}(\sigma^x\rho^z+\sigma^z\rho^x)\Psi^{\phantom\dagger}_{l}.
\label{eq:Hconduction}
\eea
The operator $\Psi_{j}$ refers to the Balian-Werthamer spinor of the conduction electrons:
\begin{equation}
\Psi_{j}=\begin{pmatrix}\psi_{j}\\
-i\sigma^{y}(\psi_{j}^{\dagger})^{T}
\end{pmatrix}
\label{Balian-Werthamer},
\end{equation}
where $\psi_{j}=(\psi_{j\uparrow} , \psi_{j\downarrow})^{T}$. 
The Pauli matrices $\sigma^{\alpha}$ and $\rho^\alpha$ act in spin space and Nambu space, respectively.  The parameters $t_{jl}=-t_{lj}$, $w_{jl}=-w_{lj}$ and $t'_{jl}=-t'_{lj}$ are defined according to the orientation of nearest- and next-nearest-neighbor bonds,  as illustrated in Fig. \ref{Haldane}: $t_{jl}=t$, $w_{jl}=w$ and $t'_{jl}=t'$ if the corresponding arrow  points from $l$ to $j$, and the opposite sign  if the arrow points from $j$  to $l$.  The Hamiltonian in Eq. \eqref{eq:Hconduction} is invariant under time reversal, $TiT^{-1}=-i$, $T\Psi_jT^{-1}=-i\sigma^y\Psi_j$, but breaks reflection  and spin-rotational symmetries. Hamiltonians similar to Eq. \eqref{eq:Hconduction} appear in the context  of  noncentrosymmetric superconductors \cite{Mineev2012,Yip2014}. Note that $t$ and $w$ alone break the sublattice inversion symmetry P: $t_{jl}\mapsto t_{lj}$, $w_{jl}\mapsto w_{lj}$, but a nonzero next-nearest-neighbor coupling $t'$ is  required to break the inversion-like  symmetry\bea
 &\textrm{P}'&\;: t_{jl}\mapsto t_{lj},\quad w_{jl}\mapsto w_{lj}, \quad t'_{jl}\mapsto t'_{lj},\nonumber\\
 &\textrm{P}'&\Psi_j (\textrm{P}')^{-1}=\left\{\begin{array}{cc}\Psi_j &\textrm{ if }j\in\textrm{sublattice A},\\
 -\Psi_j& \textrm{ if }j\in\textrm{sublattice B}.
 \end{array}\right.
\eea

It is convenient to use the electron operators $\psi_{j\uparrow}$ and $\psi_{j\downarrow}$ to define four Majorana fermions by \cite{Coleman1994}\be
\Psi_j=
\frac12\begin{pmatrix}\mathbbm{1}_2 & i\mathbbm{1}_2\\
-i\sigma^y & -\sigma^y
\end{pmatrix}\zeta_j,\label{eq:Eq_MajTrans}
\ee
where    $\mathbbm{1}_2$ is the 2$\times$2 identity matrix  and $\zeta_{j}=(\zeta_{j}^{1} , \zeta_{j}^{2} , \zeta_{j}^{3} , \zeta_{j}^{4})^{T}$, 
with   $\zeta_{j}^{\mu}$   obeying  $\{\zeta_{j}^{\mu},\zeta_{l}^{\nu}\}=2\delta_{jl}\delta^{\mu\nu}$.  More explicitly, we have\bea
\zeta^1_j&=& \psi^{\phantom\dagger}_{j\uparrow}+\psi^{ \dagger}_{j\uparrow},\nonumber\\
\zeta^2_j&=& \psi^{\phantom\dagger}_{j\downarrow}+\psi^{ \dagger}_{j\downarrow}\nonumber,\\
\zeta^3_j&=&-i (\psi^{\phantom\dagger}_{j\uparrow}-\psi^{ \dagger}_{j\uparrow}),\label{zetas}\\
\zeta^4_j&=&- i(\psi^{\phantom\dagger}_{j\downarrow}-\psi^{ \dagger}_{j\downarrow}).\nonumber 
\eea
Time reversal acts on these Majorana fermions as\bea
T:\;\zeta^1_j\mapsto \zeta^2_j,\; \zeta^2_j\mapsto -\zeta^1_j,\; \zeta^3_j\mapsto \zeta^4_j,\; \zeta^4_j\mapsto -\zeta^3_j.\label{Tzetas}
\eea
The Hamiltonian for the itinerant electrons can then be written as
\bea
H_{c}&=&i\sum_{\langle jl\rangle}\big[t_{jl}\big(\zeta_{j}^{3}\zeta_{l}^{4}+\zeta_{j}^{4}\zeta_{l}^{3}\big)+w_{jl}\big(\zeta_{j}^{3}\zeta_{l}^{3}-\zeta_{j}^{4}\zeta_{l}^{4}\big)\big]\nonumber\\
&&+i\sum_{\langle\langle jl\rangle\rangle}t'_{jl}\big(\zeta_{j}^{3}\zeta_{l}^{4}+\zeta_{j}^{4}\zeta_{l}^{3}\big).
\label{eq:Eq_CondHam}
\eea
In this form, it is clear that there are two Majorana fermions at each site, $\zeta_j^1$ and $\zeta_j^2$, which commute with $H_c$. We have defined the conduction electron Hamiltonian this way so that we can later hybridize these zero-energy Majorana modes    with the Majorana fermions in the quadrupolar spin liquid when we turn on the Kondo coupling.

\begin{figure}
\centering
\includegraphics[width=.9\columnwidth]{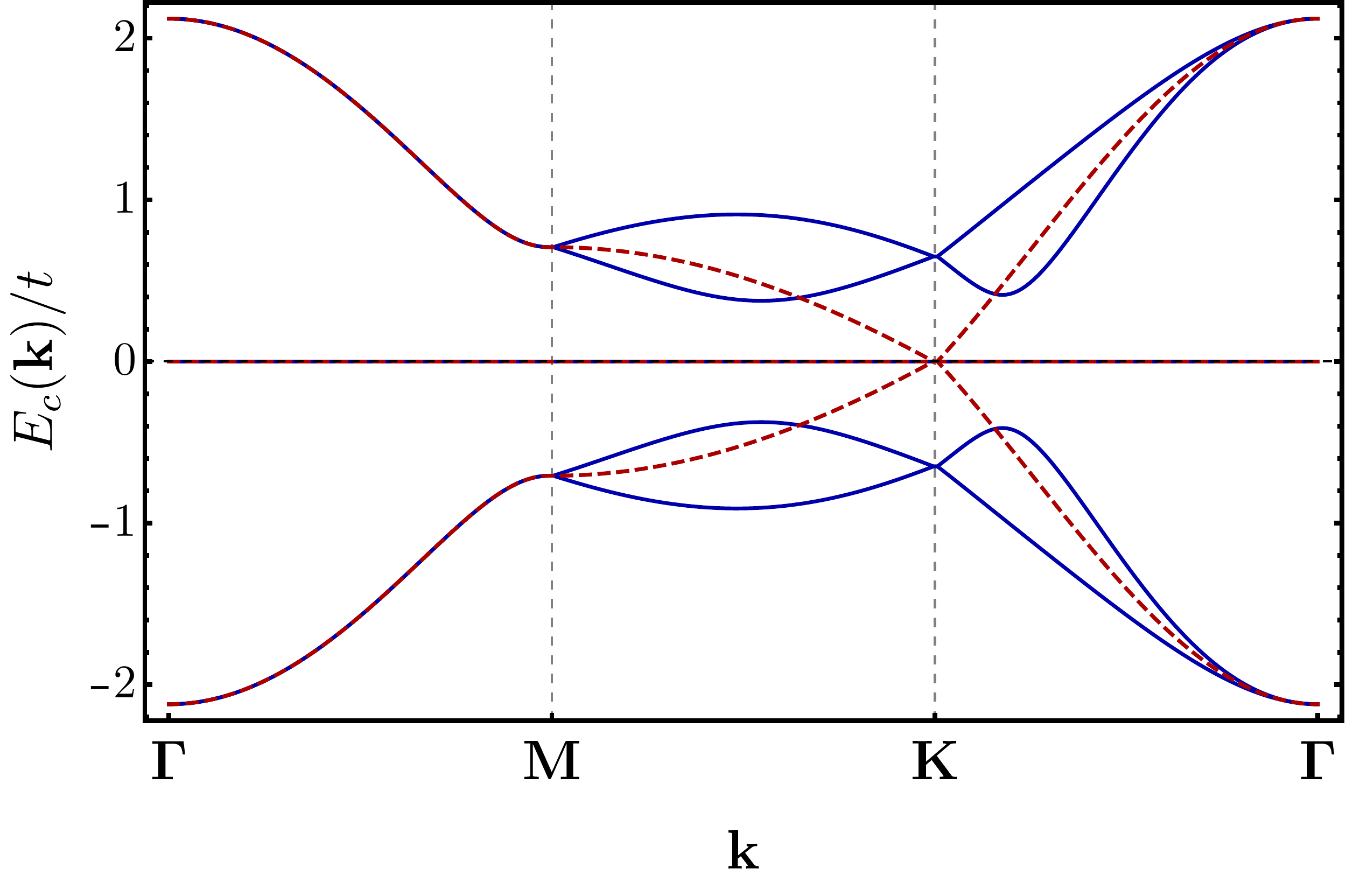}
\hspace{0.5cm}
\caption{Dispersion relation of  itinerant electrons  described by $H_c$ in Eq. (\ref{eq:Hconduction}).  Here we set $t=w$ and consider two values of the next-nearest-neighbor coupling: $t'=0$ (dashed line) and $t'=t/2$ (solid line).    The zero-energy  flat bands are related to the  $\zeta^{1}$ and $\zeta^{2}$ modes.\label{Bulk_Dispersions_Hc} }
\end{figure}

We can   diagonalize $H_c$ by taking the Fourier transform of the fermion operators. We obtain the dispersion relations%
\bea
E_{c}(\mathbf{k})&=&\pm\left[(t^{2}+w^2)|f(\mathbf{k})|^{2}+\Delta_{0}^{2}(\mathbf{k})\right.\nonumber\\
&&\qquad\left.\pm2w|\Delta_{0}(\mathbf{k})f(\mathbf{k})|\right]^{1/2},\label{Eczeta34}
\eea
where we define the functions $f(\mathbf{k})=e^{i\mathbf{k}\cdot\mathbf{n}_{1}}+e^{i\mathbf{k}\cdot\mathbf{n}_{2}}+1$ 
and $\Delta_{0}(\mathbf{k})=-2t'\sum_{\gamma=1}^{3}\sin(\mathbf{k}\cdot\mathbf{\boldsymbol{\delta}_{\gamma}})$, with  $\boldsymbol{\delta}_{1}=\mathbf{n}_{2}$,
$\boldsymbol{\delta}_{2}=-\mathbf{n}_{1}$, and $\boldsymbol{\delta}_{3}=\mathbf{n}_{1}-\mathbf{n}_{2}$.   In the  Majorana fermion basis, these bands are associated with $\zeta^3$ and $\zeta^4$, whereas the $\zeta^1$ and $\zeta^2$ fermions appear in  the spectrum as zero-energy flat bands, see  Fig. \ref{Bulk_Dispersions_Hc}. The gap in the $\zeta^{3,4}$ sector is of order $|t'|$ if $|t'|\ll |t|,|w|$. We are mainly interested in the regime $|t'|\sim |t| \sim |w| \gg |K|,|\lambda|$, in which  this gap is much larger than the interaction energy scale for the spin liquid. In this limit, we can project out the $\zeta^{3,4}$ sector and the low-energy physics is governed by the  coupling between the $\zeta^{1,2}$ modes of the conduction electrons and the localized spins.

\subsection{Octupolar Kondo coupling\label{sec:Kondo}}

We now look for an on-site interaction that couples electrons to local moments and preserves time reversal symmetry. First, we note that the projection of the electron spin operator onto the sector of $\zeta^{1,2}$ Majorana fermions is given by 
\bea
\mc P_{12} \Psi^\dagger_j\boldsymbol\sigma\Psi^{\phantom\dagger}_j\mc P_{12}&=& \frac12\Psi^\dagger_j(\sigma^y-\rho^y)\Psi^{\phantom\dagger}_j  \hat {\mb y} \nonumber\\
&=&-i\zeta^1_j\zeta^2_j \hat {\mb y},\label{projectspin}
\eea
where $\mc P_{12}$ is the projection operator. Note that only the $y$ component of the spin operator remains after the projection; this is possible because the spin-rotational symmetry is   broken in the   Hamiltonian in Eq. (\ref{eq:Hconduction}). If the superconducting gap in the $\zeta^{3,4}$ sector far exceeds all other energy scales, the operator in Eq. (\ref{projectspin}) is the only single-site electron operator active at low energies. We then consider the Kondo-like coupling  
\begin{equation}
H_{K}= \frac{J_{K}}2\sum_{j} \Psi_{j}^{\dagger}(\sigma^y-\rho^y)\Psi^{\phantom\dagger}_{j} T_{j}^{xyz}.
\label{eq:Octupolar_Kondocoupling}
\end{equation}
This interaction involves the octupole moment $T_j^{xyz}\propto \tau_j^y$ of the localized spins.In terms of Majorana fermions, we have
\begin{equation}
H_{K}=J_{K}\sum_{j}\zeta_{j}^{1}\zeta_{j}^{2}\theta_{j}^{x}\theta_{j}^{z}.\label{KondoMajorana}
\end{equation}
Note that the condition that the Kondo-like coupling must respect the conservation of the  bond operators $\hat{u}_{\langle jl\rangle_\gamma}$ prevents us from coupling the projection of the electron spin operator in Eq. (\ref{projectspin}) to the dipole moments $\mb J_j$ of the localized spins, see Table \ref{tab:multipoles}. We thus expect  the coupling in Eq. (\ref{eq:Octupolar_Kondocoupling}) to arise from more microscopic models as the leading interaction between itinerant and localized electrons at   energy scales far below the superconducting gap shown in Fig. \ref{Bulk_Dispersions_Hc} and  the gap for changing the  flux configuration in the spin liquid. 

To solve the Hamiltonian in Eq. (\ref{eq:KitaevKondo}), we note that we can pair the $\theta_{j}^{z}$ Majorana fermion with either $\zeta_{j}^{1}$ or $\zeta_{j}^{2}$ to define the $\mathbbm Z_2$ operators\be
\hat v_j=-i\zeta_j^{\mu_j}\theta_j^z,\label{Z2vs}
\ee 
where $\mu_j\in\{1,2\}$ can be chosen independently at each site. It is straightforward  to verify that these operators commute not only with one another, but also with the bond operators $\hat{u}_{\langle ij\rangle_\gamma}$ of the spin liquid and with the total Hamiltonian in Eq. (\ref{eq:KitaevKondo}). Thus, $\hat v_j$ are conserved quantities and we can replace them  by the eigenvalues $\pm1$ to obtain a quadratic Hamiltonian in the remaining Majorana fermions. Remarkably,  a finite expectation value of the   $\hat v_j$ operators implies a hybridization between physical Majorana fermions defined from the conduction electrons and emergent Majorana fermions in the spin  liquid. This  type of hybridization has appeared in the literature as an order parameter for odd-frequency pairing  in heavy-fermion superconductors \cite{Coleman1993,Coleman1994}.

Since time reversal exchanges $\zeta^1$ and $\zeta^2$, see Eq. (\ref{Tzetas}), the choice of $\mu_j=1,2$ for each  $\hat v_j$  breaks time reversal symmetry  spontaneously. Thus, the   Kondo coupling to the octupole moment of the  spin liquid  induces   unconventional, time-reversal-symmetry-breaking superconductivity. To discuss the possible superconducting states, we must first assign   values to the  $\mathbbm Z_2$ variables.  We consider two states that respect the translational symmetry of the honeycomb lattice: the uniform  or ``ferro'' (F) configuration   \be
\hat v^{\textrm{F}}_j=-i\zeta_j^1\theta_j^z=1\qquad \forall j, 
\ee
and the staggered or ``antiferro'' (AF) configuration\be
\hat v^{\textrm{AF}}_j=\left\{ \begin{array}{cc}
-i\zeta_j^1\theta_j^z=1,& \textrm{if } j\in \textrm{ sublattice A}, \\
-i\zeta_j^2\theta_j^z=1,& \textrm{if } j\in \textrm{ sublattice B}.
\end{array}  \right.
\ee
Changing the sign of $\hat v_j$ does not affect any physical observables, due to the gauge symmetry $(\eta_j^\alpha,\theta_j^\alpha)\mapsto (-\eta_j^\alpha,-\theta_j^\alpha)$ of the Majorana fermion representation  in the spin liquid sector. The Kondo coupling for F and AF configurations becomes, respectively, \bea
H_{K}^{\textrm{F}}&=&iJ_{K}\sum_{j}\zeta_{j}^{2}\theta_{j}^{x},\label{KondoF}\\
H_{K}^{\text{AF}}&=&iJ_{K}\sum_{j\in\textrm{A}}\zeta_{j}^{2}\theta_{j}^{x}-iJ_{K}\sum_{j\in\textrm{B}}\zeta_{j}^{1}\theta_{j}^{x}.\label{KondoAF}
\eea

\begin{figure}
\centering
\includegraphics[width=.9\columnwidth]{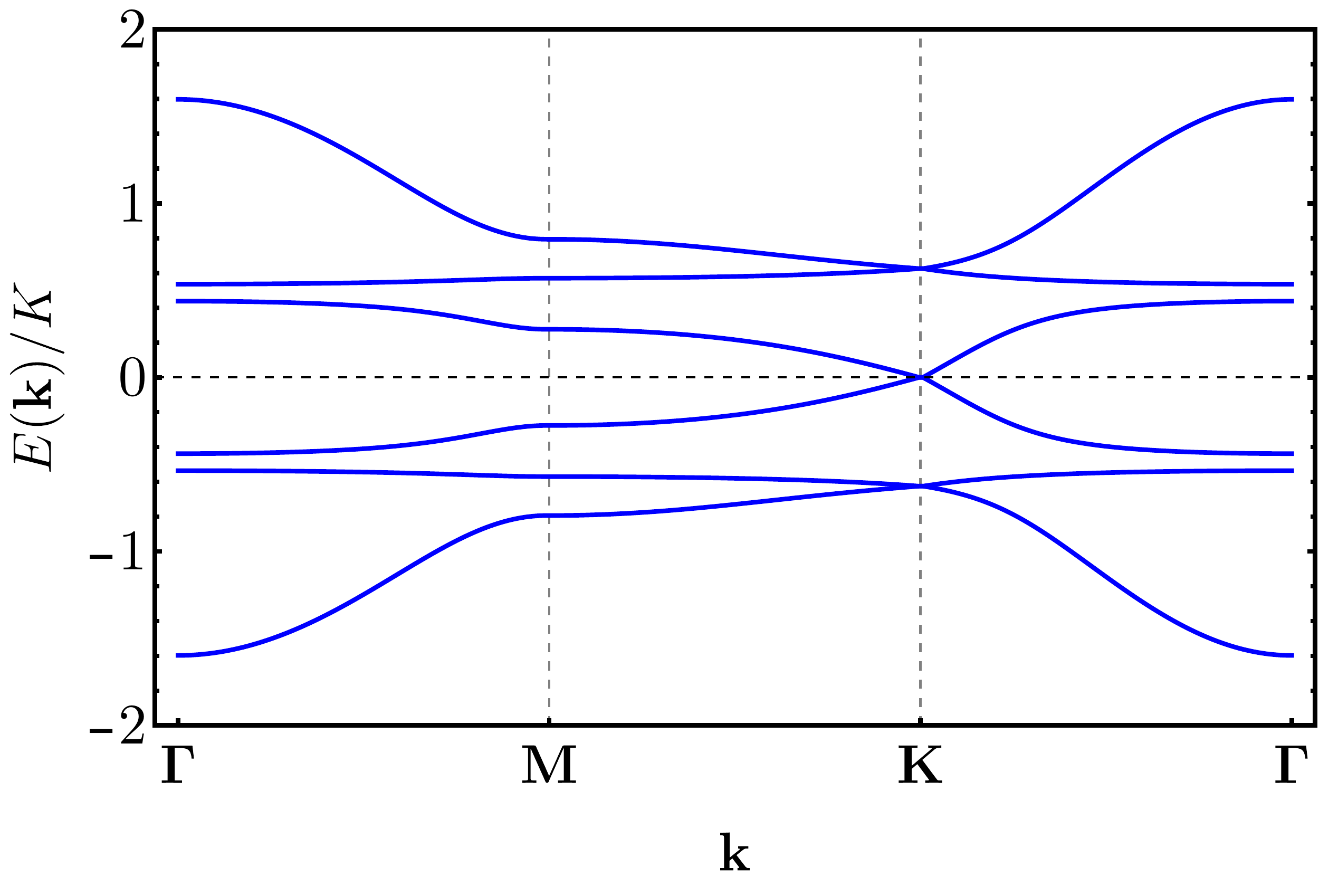}\\ 
\caption{  Dispersion relation of   Majorana fermions in both F and AF configurations of  the coupled system.  Here we set $\lambda/K=0.75$ and $J_{K}/K=1$.  The  bands associated with  the decoupled $\zeta^{3}$ and $\zeta^{4}$ Majorana fermions appear at much higher energies and are not shown.   
\label{Ground_State}}
\end{figure}

In order to obtain the energy spectrum with finite Kondo interaction, we define the   spinors
\begin{equation}
\Upsilon_{\text{F}}(\mathbf{k})=\begin{pmatrix}\theta^x_{\textrm{A}}(\mathbf{k})\\
\theta^x_{\textrm{B}}(\mathbf{k})\\
\theta^y_{\textrm{A}}(\mathbf{k})\\
\theta^y_{\textrm{B}}(\mathbf{k})\\
\zeta_{\textrm{A}}^{2}(\mathbf{k})\\
\zeta_{\textrm{B}}^{2}(\mathbf{k})
\end{pmatrix},
\quad
\Upsilon_{\text{AF}}(\mathbf{k})=\begin{pmatrix}\theta^x_{\textrm{A}}(\mathbf{k})\\
\theta^x_{\textrm{B}}(\mathbf{k})\\
\theta^y_{\textrm{A}}(\mathbf{k})\\
\theta^y_{\textrm{B}}(\mathbf{k})\\
\zeta_{\textrm{A}}^{2}(\mathbf{k})\\
-\zeta_{\textrm{B}}^{1}(\mathbf{k})
\end{pmatrix},
\end{equation}
which contain  the Majorana fermions that  acquire a dispersion at low energies. The total  Hamiltonian   has the form  
\begin{equation}
\mc P_{12}H^{\textrm{F}/\textrm{AF}}\mc P_{12}=\sum_{\mathbf{k}}\Upsilon_{\textrm{F}/\textrm{AF}}^{\dagger}(\mb k)\mc H(\mb k)\Upsilon^{\phantom\dagger}_{\textrm{F}/\textrm{AF}}(\mathbf{k}),
\end{equation}
where  the antisymmetric matrix $\mc H(\mb k)$ is the same for both uniform and staggered configurations:
\be
\mc H(\mb k)=\left(\begin{array}{cc}
\mc H_s(\mb k)&V^\dagger\\
V& \mathbb 0_2
\end{array}
\right),
\ee
with $\mathbb 0_2$ the  2$\times$2 null matrix, 
\be
\mc H_s(\mb k)=\left(\begin{array}{cccc}
0&0&i\lambda&0\\
0&0&0&i\lambda\\
-i\lambda&0&0&ig(\mb k)\\
0&-i\lambda&-ig^*(\mb k)&0
\end{array}
\right),\label{Hsk}
\ee
and \be
V=\left(\begin{array}{cccc}
-iJ_K&0&0&0\\
0&-iJ_K&0&0
\end{array}\right).
\ee
Thus,   F and AF states share the same energy spectrum illustrated  in Fig. \ref{Ground_State}.  We see that the Majorana fermions that appear in the Kondo coupling Eqs. (\ref{KondoF}) and  (\ref{KondoAF}) give rise to  a gapless band structure with a Dirac node at the K point. This represents the dispersion of  the  Bogoliubov quasiparticles in this nodal superconductor. There are no zero-energy flat bands left.  The dispersion relations for the decoupled $\zeta^{3,4}$ modes are given by Eq. (\ref{Eczeta34}) and appear at much higher energies provided that  $|t|\sim |t'|\sim |w|\gg |K|, |\lambda|, |J_K|$.

Since we obtain the  same spectrum for  F and AF configurations, these two states  have exactly the same ground state energy. This degeneracy can be traced back to a local SO(2) symmetry of the exactly solvable model. We can regard $\zeta^1_j$ and $\zeta^2_j$ as the real and imaginary parts of a complex fermion defined  at each site. The choice of real and imaginary parts   can be parametrized by a rotation in the complex plane, i.e., the phase of the complex fermion. This choice  is   arbitrary and  can be made locally at each site because   these Majorana modes only appear in the Kondo coupling Eq. (\ref{KondoMajorana}). As we shall discuss in Sec. \ref{sec:perturb}, the degeneracy between F and AF states is lifted once we add  integrability-breaking    perturbations that couple  $\zeta^1$ and $\zeta^2$ at different sites and  remove the local SO(2) symmetry.

Interestingly, despite the spontaneous  time reversal symmetry breaking, the local magnetization vanishes exactly in the ground state of the exactly solvable model: \be
\langle \Psi_j^\dagger \boldsymbol\sigma\Psi^{\phantom\dagger}_j\rangle=0,\qquad \langle \mb J_j\rangle=0.\ee
The reason is that  both the electron and local moment  spin operators contain Majorana fermions absorbed into the $\mathbbm Z_2$ operators, whose action on a given eigenstate changes the sector of  $\mathbbm Z_2$ fluxes.  Local time-reversal-odd operators that acquire a nonzero expectation value must involve a product of electron and local-moment operators, for instance, \bea
\langle O_j^{x^2-y^2}\Psi^\dagger_j \sigma^y\Psi^{\phantom\dagger}_j\rangle&\sim&  \langle \theta_j^y\theta_j^z\zeta_j^1\zeta_j^2\rangle\nonumber\\
&\sim&  \langle \zeta_j^1\theta_j^z\rangle\langle \zeta_j^2\theta_j^y\rangle  \neq0 ,
\eea 
where the factor $\langle \zeta_j^2\theta_j^y\rangle$ is nonzero because $\lambda\neq0$ mixes $\theta^y_j$ with $\theta^x_j$ and the operator  $   \zeta_j^2\theta_j^x $ appears in the Kondo coupling in Eq. (\ref{KondoF}). This provides an example of a composite order parameter \cite{Emery1992,Emery1993,Abrahams1995,Zachar1996,Hoshino2011,Hoshino2013,Chandra2013,Dyke2019,Balatsky2019} or vestigial order \cite{Fernandes2019}.

\section{Effective action for electrons in the coupled system \label{sec:Instabilities-induced-by}}

In this section, we investigate the effect of the octupolar Kondo coupling on the superconducting properties of the coupled system. Since the Hamiltonian becomes quadratic after fixing the values of the $\mathbbm Z_2$ variables, we can  integrate out the Majorana fermions of the quadrupolar spin liquid to derive an exact effective action for the conduction electrons. We are mainly interested in the   pairing amplitudes generated by the Kondo coupling which manifest the breaking of  time reversal symmetry. We find that, while the uniform  and staggered  configurations are   degenerate within the exactly solvable  model, they produce different superconducting order parameters because the physical  Majorana fermions $\zeta^1$ and $\zeta^2$ are associated with different electronic spin states, see Eq. (\ref{zetas}). More details of the calculations are given in Appendix \ref{App1}.  

\subsection{Superconducting state in the F  configuration\label{sec:SCF}}

For the uniform state with the Kondo interaction $H_{K}^{\textrm{F}}$ in Eq. (\ref{KondoF}), the effective action obtained after integrating out the Majorana fermions $\theta^x$ and  $\theta^y$ has the form
\begin{equation}
\mathcal{S}_{\textrm{F}}=\sum_{\mathbf{k},\omega_n}   \Psi^\dagger(\mathbf{k},i\omega_n) [i\omega_n-\mathcal{H}_c(\mathbf{k})-\Sigma_{\textrm{F}}(\mathbf{k},i\omega_n)]\Psi(\mb k,i\omega_n),\label{action}
\end{equation}
where $\omega_n$ are Matsubara frequencies and   $\Psi(\mb k,i\omega_n)$ is the   Balian-Werthamer spinor in momentum-frequency space:
\be
\Psi(\mb k,i\omega_n)=\begin{pmatrix} \Psi_{\textrm{A}}(\mathbf{k},i\omega_n) \\ \Psi_{\textrm{B}}(\mathbf{k},i\omega_n) \end{pmatrix},
\ee
with
\bea
\Psi_{\textrm{A/B}}(\mathbf{k},i\omega_n)&=\begin{pmatrix} \psi_{\textrm{A/B}}(\mathbf{k},i\omega_n) \\ -i\sigma^y[\psi^\dagger_{\textrm{A/B}}(-\mathbf{k},-i\omega_n)]^T \end{pmatrix}.\label{Eq_mom_freq_BW}
\eea
Here $\mc H_c(\mb k)$ is the matrix obtained by Fourier transforming $H_c$ in Eq. (\ref{eq:Hconduction}), see Appendix \ref{App1}. The self-energy $\Sigma_{\textrm{F}}(\mathbf{k},i\omega_n)$ is due to the hybridization  of the conduction electrons with the Majorana fermions of the spin liquid and is of order $J_K^2$.

The effective action in Eq. (\ref{action})   contains three terms: a normal (N), a superconducting (SC), and a resonant-exchange (RE) part. The N and RE parts are  written explicitly in Appendix \ref{App1}. Here we discuss only the contribution from the Kondo coupling to the SC part, which has the form  
\bea
\delta\mathcal{S}^\text{SC}_{\textrm{F}}&=& \sum_{ 	\mathbf{k},\omega_n}\sum_{b,b'}\big[\psi^T_b(-\mathbf{k},-i\omega_n)\sigma^y\Delta_{bb'}^\text{SC}(\mathbf{k},i\omega_n)\psi_{b'}(\mathbf{k},i\omega_n)\nonumber\\
&&\qquad\quad+\text{H.c.}\big], 
\label{FK_SC_action}
\eea
where $b,b'$  are sublattice indices. The induced pairing functions are given by
\begin{align}
\Delta_{\textrm{AA}}^\text{SC}(\mathbf{k},i\omega_n)&=\frac{J^2_K}2\frac{\omega_n(\omega^2_n+|g(\mathbf{k})|^2+\lambda^2) \sigma^+}{(\omega^2_n+\lambda^2)^2+|g(\mathbf{k})|^2\omega^2_n},\nonumber\\
\Delta_{\textrm{AB}}^\text{SC}(\mathbf{k},i\omega_n)&=-\frac{J^2_K}2\frac{\lambda^2g(\mathbf{k})  \sigma^+}{(\omega^2_n+\lambda^2)^2+|g(\mathbf{k})|^2\omega^2_n},\nonumber\\
\Delta_{\textrm{BA}}^\text{SC}(\mathbf{k},i\omega_n)&=\frac{J^2_K}2\frac{\lambda^2g^*(\mathbf{k}) \sigma^+}{(\omega^2_n+\lambda^2)^2+|g(\mathbf{k})|^2\omega^2_n},\nonumber\\
\Delta_{\textrm{BB}}^\text{SC}(\mathbf{k},i\omega_n)&=\frac{J^2_K}2\frac{\omega_n(\omega^2_n+|g(\mathbf{k})|^2+\lambda^2) \sigma^+}{(\omega^2_n+\lambda^2)^2+|g(\mathbf{k})|^2\omega^2_n},
\label{FK_pairing}
\end{align}
where $\sigma^{\pm}=(\sigma^x\pm i\sigma^y)/2$. Therefore, the Kondo coupling gives rise to triplet pairing with both even- and odd-frequency amplitudes. Exactly at the K point, where $g(\mb k)$ vanishes for $K_\gamma=K$,   the  AB and BA  components vanish, whereas the  AA and BB components scale linearly with frequency, $\Delta_{bb}^\text{SC}\sim (J_K/\lambda)^2\omega_n\sigma^+$ for $|\omega_n|\ll |\lambda|$.  This result bears  a close resemblance to  mean-field theories for related Kondo lattice models \cite{Coleman1994,Tsvelik2017,Balatsky2019}. 

\subsection{Superconducting state in the AF configuration}

We repeat the procedure of Sec. \ref{sec:SCF} for the staggered state with the Kondo interaction given in Eq. (\ref{KondoAF}).   The effective action in this case also exhibits    N, SC and RE contributions. The superconducting part generated by  the Kondo coupling in the AF case reads
\bea
\delta {\mathcal{S}}^\text{SC}_{\textrm{AF}}&=&\sum_{ 	\mathbf{k},\omega_n}\sum_{b,b'}\big[\psi^T_b(-\mathbf{k},-i\omega_n)\sigma^y\tilde\Delta_{bb'}^\text{SC}(\mathbf{k},i\omega_n)\psi_{b'}(\mathbf{k},i\omega_n)\nonumber\\
&&\qquad\quad+\text{H.c.}\big], 
\label{AFK_SC_action}
\eea
where
\begin{align}
\tilde{\Delta}_{\text{AA}}^\text{SC}(\mathbf{k},i\omega_n)&=\frac{J^2_K}2\frac{\omega_n(\omega^2_n+|g(\mathbf{k})|^2+\lambda^2) \sigma^+}{(\omega^2_n+\lambda^2)^2+|g(\mathbf{k})|^2\omega^2_n},\nonumber\\
\tilde{\Delta}_{\text{AB}}^\text{SC}(\mathbf{k},i\omega_n)&=\frac{J^2_K}4\frac{\lambda^2g(\mathbf{k}) (\mathbb1_2+\sigma^z)}{(\omega^2_n+\lambda^2)^2+|g(\mathbf{k})|^2\omega^2_n},\nonumber\\
\tilde{\Delta}_{\text{BA}}^\text{SC}(\mathbf{k},i\omega_n)&=\frac{J^2_K}4\frac{\lambda^2g^*(\mathbf{k}) (\mathbb 1_2-\sigma^z)}{(\omega^2_n+\lambda^2)^2+|g(\mathbf{k})|^2\omega^2_n},\nonumber\\
\tilde{\Delta}_{\text{BB}}^\text{SC}(\mathbf{k},i\omega_n)&=-\frac{J^2_K}2\frac{\omega_n(\omega^2_n+|g(\mathbf{k})|^2+\lambda^2) \sigma^-}{(\omega^2_n+\lambda^2)^2+|g(\mathbf{k})|^2\omega^2_n}.
\label{AFK_pairing}
\end{align}
These pairing  functions also involve a mixture of   even- and odd-frequency pairing. However, they are markedly different from those in the uniform configuration, since now  the AB and BA    components exhibit a superposition of singlet and triplet pairings. Moreover, the  AA and BB components differ in the spin  dependence of the triplet pairing.  At the K point, $\Delta_{\textrm{AA}}^\text{SC}\sim (J_K/\lambda)^2\omega_n\sigma^+$ for the A sublattice, but  $\Delta_{\textrm{BB}}^\text{SC}\sim (J_K/\lambda)^2\omega_n\sigma^-$ for the B sublattice.

\section{Integrability-breaking perturbations  \label{subsec:Perturbations}\label{sec:perturb}}


In this section, we go beyond the exactly solvable model to examine whether    perturbations  may lift the degeneracy between the  F and AF configurations. Rather than discuss all symmetry-allowed perturbations, we focus on  the effects of 
 additional   quadratic terms  in  the conduction  electron Hamiltonian that     involve the Majorana fermions $\zeta^{1}$ and $\zeta^{2}$. We consider
\bea
\delta H_c&=&i\sum_{\langle jl\rangle}[\delta t_{jl}(\zeta^1_j\zeta^2_l+\zeta^2_j\zeta^1_l)+\delta w_{jl}(\zeta^1_j\zeta^1_l-\zeta^2_j\zeta^2_l)]\nonumber\\ 
&&+i\sum_{\langle\langle jl\rangle\rangle}[\delta t'_{jl}(\zeta^1_j\zeta^2_l+\zeta^2_j\zeta^1_l)+\delta w'_{jl}(\zeta^1_j\zeta^1_l-\zeta^2_j\zeta^2_l)],\nonumber\\
&&
\label{perturbationNN}
\eea
where we fix the values of $\delta t_{jl}=\pm \delta t$, $\delta w_{jl}=\pm \delta w$, $\delta t'_{jl}=\pm \delta t'$ and $\delta w'_{jl}=\pm \delta w'$ according to the   orientation of the nearest and next-nearest-neighbor links as explained in Sec. \ref{Time-reversal_symmetric_SC}. In terms  of the   Balian-Werthamer spinor [see Eq.  (\ref{Balian-Werthamer})],  $\delta H_{c}$ assumes the form 
\bea
\delta H_{c}& =&\sum_{\langle jl\rangle} \Psi^\dagger_jM_{jl}\Psi^{\phantom\dagger}_l+\sum_{\langle\langle jl\rangle\rangle} \Psi^\dagger_jM_{jl}'\Psi^{\phantom\dagger}_l,
\eea
with
\bea
\hspace{-.3cm}M_{jl}&=&i\delta t_{jl}(\sigma^x\rho^z-\sigma^z\rho^x)   +i\delta w_{jl}(\sigma^z\rho^z+\sigma^x\rho^x),\nonumber\\
\hspace{-.3cm}M'_{jl}&=&i\delta t'_{jl}(\sigma^x\rho^z-\sigma^z\rho^x)   +i\delta w'_{jl}(\sigma^z\rho^z+\sigma^x\rho^x).
\eea
 %
%
Since the $\mathbbm Z_2$ operators $\hat v_j$ defined in Eq. (\ref{Z2vs}) do not commute with $\delta H_c$, this perturbation breaks the integrability of the model.  Here we shall assume that $|\delta t|,|\delta w|,|\delta t'|,|\delta w'|\ll |K|,|\lambda|,|J_K|$ so that the $\mathbbm Z_2$ variables are still good order parameters with expectation value $\langle\hat v_j \rangle\approx 1$.  One consequence of the perturbation is that the Majorana fermions absorbed into $\hat v_j$ must acquire a small dispersion. Nevertheless, as long as the latter remain gapped, this should not affect qualitative properties of the  low-energy spectrum discussed in the following.

\begin{figure}
	\centering
	\includegraphics[scale=0.3]{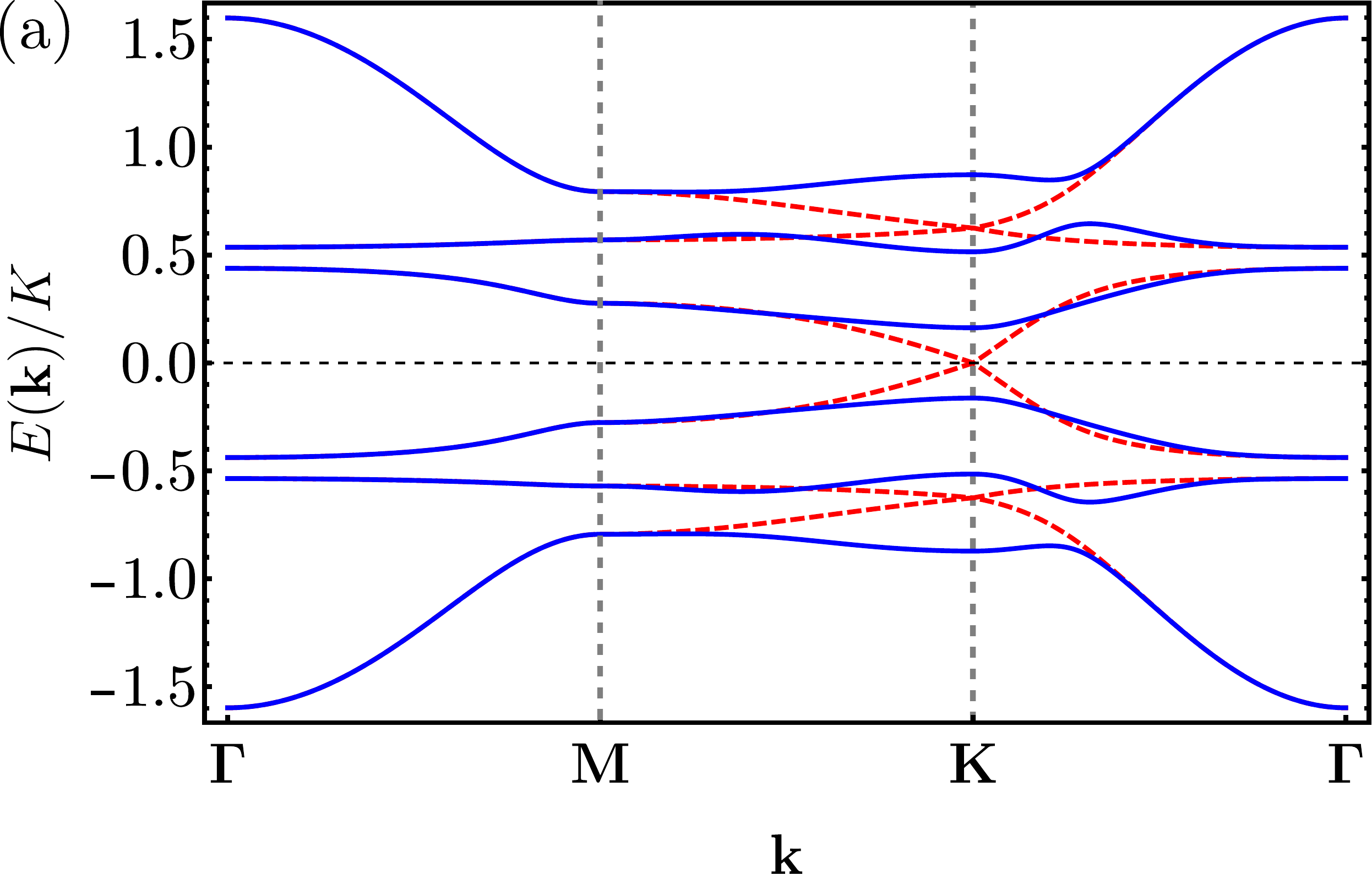}\hspace{0.3cm}
	\includegraphics[scale=0.3]{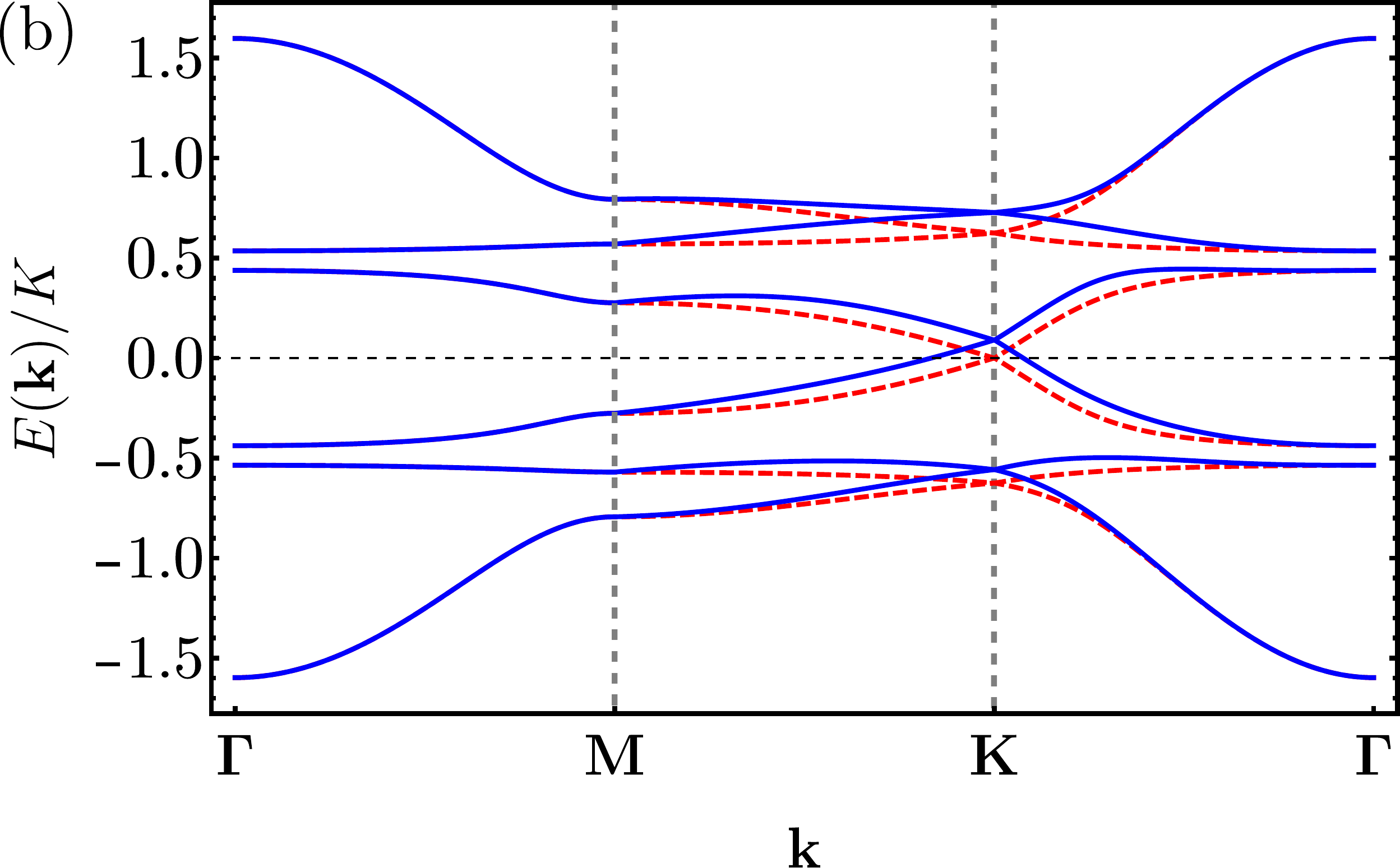}\hspace{0.3cm}
	\caption{\small{Low-energy bands  of the coupled system including the integrability-breaking  perturbations in the (a) F configuration and (b) AF configuration. Here we use $\lambda/K=0.75$, $J_{K}/K=1$ and  $\delta t=\delta w=0$. The    solid   lines show the result for $\delta w'/K=0.2$, which can be compared with the dispersion   for the exactly solvable model  ($\delta w'=0$) given by the    dashed lines.  For the F state, a small    $\delta w'$ opens   a gap at the  K point.  For the AF state,  it induces the formation of a Bogoliubov Fermi surface.}}
	\label{Bulk_Dispersions_withPertubation}
\end{figure}

In the spirit of  first-order  perturbation theory, we   project   $\delta H_{c}$ onto the low-energy subspace where the Majorana fermions contained in $\hat v_j$ are gapped out and cannot be excited.   For the F configuration, this rules out the terms   involving  $\zeta^1$ altogether. As a result,   we obtain the projected perturbation 
\begin{equation}\label{Eq_FPert}
\delta H^{\text{F}}_{c}=-i\sum_{\langle jl\rangle}\delta w_{jl}\zeta^2_j\zeta^2_l-i\sum_{\langle\langle jl\rangle\rangle}\delta w'_{jl}\zeta^2_j\zeta^2_l.
\end{equation}
In contrast, in the AF configuration   the fermionic excitations   related to $\zeta^1$ become gapped only in sublattice A, while those related to $\zeta^2$ are gapped in sublattice B. Consequently, the   projection of $\delta H_c$ for the AF state yields
\bea
\delta H^{\text{AF}}_{c}&=&
i\sum_{\substack{\langle jl\rangle \\ j\in \text{B}}
}\delta t_{jl}\zeta^1_j\zeta^2_l
+i\sum_{\substack{\langle\langle jl\rangle\rangle \\ j\in\text{B}}}
\delta w'_{jl} \zeta^1_j\zeta^1_l
\nonumber \\
&&-i\sum_{\substack{\langle\langle jl\rangle\rangle\\ j \in\text{A}}}\delta w'_{jl}\zeta^2_j\zeta^2_l.
\label{Eq_AFPert}
\eea
Note that both    $\delta H^{\text{F}}_{c}$  and $\delta H^{\text{AF}}_{c}$ remove the  local  SO(2) symmetry that exchanges $\zeta^1_j$  and $\zeta^2_j$, see Sec. \ref{sec:Kondo}.  

We add the terms in  Eqs. \eqref{Eq_FPert} and \eqref{Eq_AFPert} to the Hamiltonian for the coupled system in the  F and AF configurations, respectively, and recalculate the spectrum by taking the Fourier transform of the Majorana fermion operators.    
 The   dispersion relations    are shown in Fig. \ref{Bulk_Dispersions_withPertubation}.  We find that the degeneracy between F and AF states is lifted by a finite next-nearest-neighbor coupling $\delta w'$. The latter  produces a gap in the excitation spectrum for the F state, while for the AF state it turns the Dirac point into a Bogoliubov Fermi surface \cite{Brydon2018}.  The ground state energies of the two configurations  are now clearly different, and either state can have  lower energy depending on the values of $\delta  t$, $\delta w$ and $\delta  w'$.

\begin{figure}[t]
	\centering \includegraphics[width=.9\columnwidth]{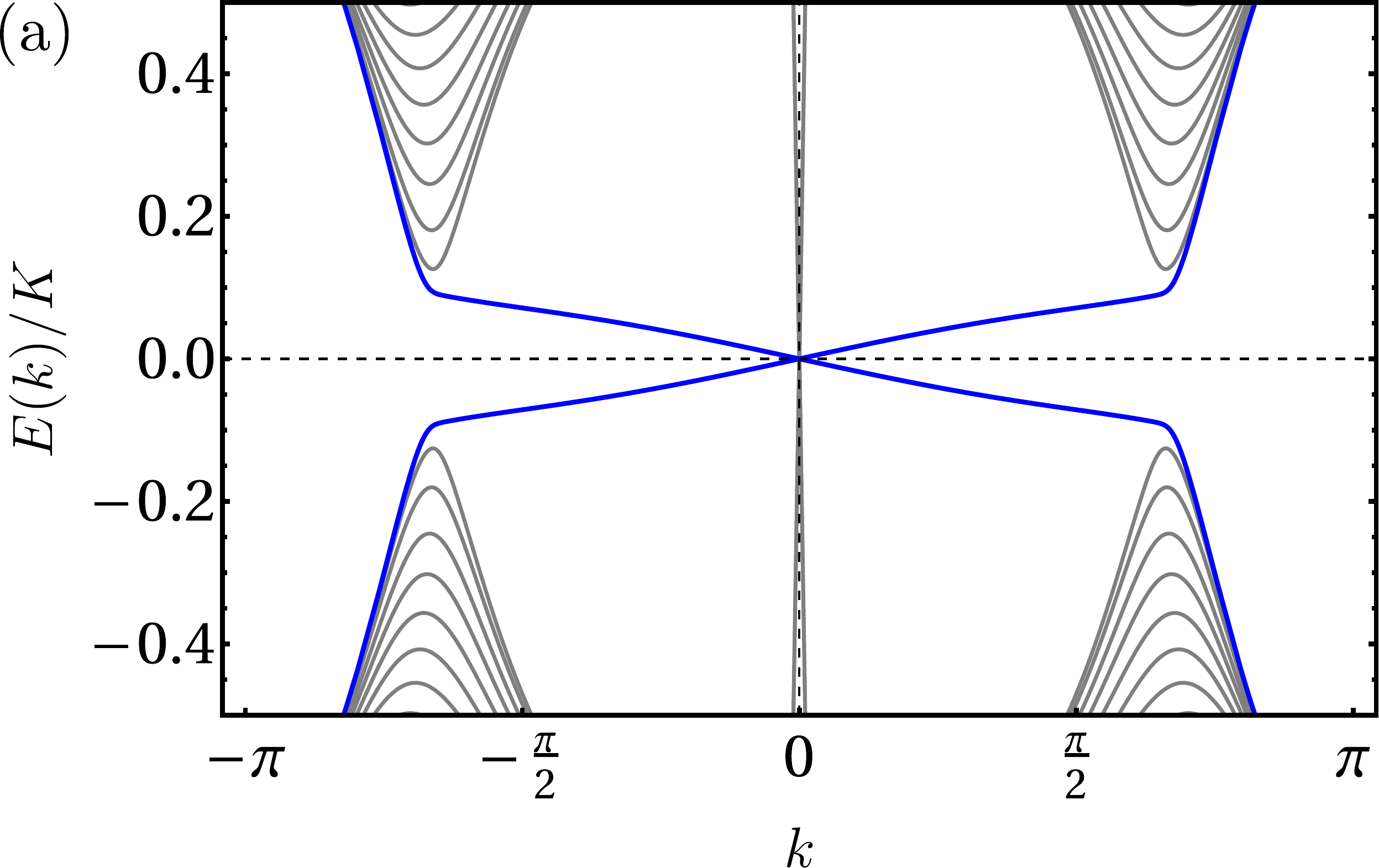}\hspace{0.01cm}
	\includegraphics[width=.9\columnwidth]{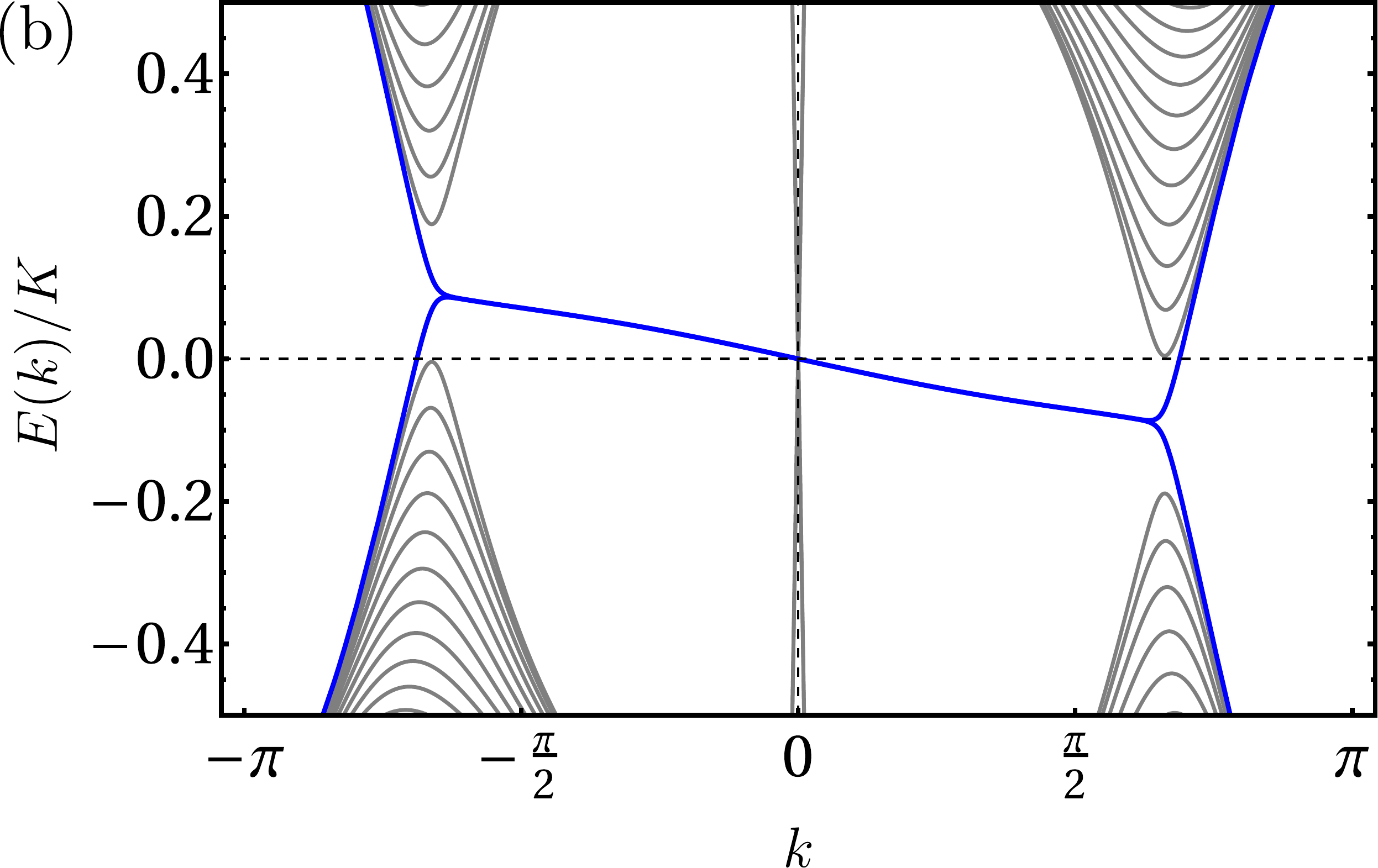}\hspace{0.01cm}
	\caption{\small{Band structure of the perturbed model   for the coupled    system with open  boundary conditions    in the $y$ direction and width $W=60$ unit cells.  Here we set $J_K/K=1$, $\lambda/K=0.75$, and $\delta t/K=\delta w/K=\delta w'/K=0.1$. Panel (a)    shows the spectrum for the F state.  The blue lines highlight the gapless edge states associated with  the low-energy  Majorana fermions.  We also show the pair of edge states associated with the $\zeta^{3,4}$ sector of the conduction electrons, whose bulk excitations appear at much higher energies. The parameters in this sector are set to $t/K=w/K=10$ and $t'/K=5$.  Panel (b) shows  the spectrum for  the AF state. In this case we find antichiral edge states.  }}
	\label{Edge_Dispersions}
\end{figure}
 
We also investigated  the presence of edge states for the perturbed superconducting states on a strip geometry. The results are presented in Fig. \ref{Edge_Dispersions}. For both F and AF states, there exist pairs of gapless edge mode due to  the nontrivial topological nature of the phase. The perturbed F state is a gapped superconductor with counter-propagating chiral edge modes localized at opposite edges of the strip. On the other hand, the AF state has edge states that propagate in the same direction, and whose equilibrium current is  compensated by that of the gapless bulk modes. This is a superconducting version of the antichiral edge states discussed in Ref. \cite{Franz2018}.    Note that both types of band structures shown in Fig.  \ref{Edge_Dispersions} are only possible once time reversal symmetry is broken.

\section{Conclusions\label{sec:Discussions-and-conclusions}}

We proposed an exactly solvable model  for interacting  $j_{\textrm{eff}}=\frac32$  local moments on the honeycomb lattice. Our proposal is guided by   symmetry properties and by a Majorana fermion representation of the multipole operators. We first analyzed  a time-reversal-invariant spin model that includes bond-dependent quadrupole-quadrupole interactions and a single-ion anisotropy term. To obtain a solvable spin Hamiltonian  with no zero-energy flat bands, we added terms that break time reversal symmetry explicitly and found a gapped chiral spin liquid. 

We also investigated the coupling of the time-reversal-invariant quadrupolar spin liquid to a superconductor, with the goal of  constructing  an exactly solvable model in which the Majorana fermions in the spin liquid hybridize with itinerant electrons. The conserved $\mathbbm Z_2$ variables defined in the octupolar Kondo coupling are related to the order parameter  for odd-frequency  pairing  in  heavy-fermion superconductors \cite{Coleman1993,Coleman1994}. Indeed, we find that  this Kondo coupling  breaks time reversal symmetry spontaneously and generates odd-frequency pairing in the effective action for the conduction electrons. The result within the exactly solvable model is a gapless time-reversal-symmetry-breaking superconductor. Perturbing the model with integrability breaking terms, we obtained either a gapped chiral superconductor or  a Bogoliubov  Fermi surface state, both of which exhibit topologically protected (chiral or antichiral) edge states.

Our results illustrate  the possibility of inducing new topological phases by coupling two subsystems which may or may not have topological properties by themselves. This observation is in line with the bulk topological proximity effect introduced in Refs. \cite{Hsieh2016, Hsiehe2017} and with the proposal of  topological superconductivity  in the Kondo-Kitaev model \cite{Choi2018}.  Also noteworthy is the recent
experimental evidence for odd-frequency superconductivity at the interface between a topological insulator and a conventional superconductor \cite{Krieger2020}.  While in this work we have focused on analyzing  the properties of an exactly solvable toy model, an interesting open question is whether  one could tune  more realistic  effective Hamiltonians for  spin-orbit-coupled materials  to the vicinity of this integrable point. 
 Besides $j_{\text{eff}}=\frac32$ systems, the spin-orbital physics   discussed  here   could be relevant to correlated Moir\'e systems   \cite{Cao2018, Po2018} and heterostructures \cite{Novoselov2016} with orbital/valley degrees of freedom.


\begin{acknowledgments}
We thank   M. M. de Oliveira, V. Quito,  and S. Trebst for stimulating discussions. We are grateful to F. Ramos for checking the ground state energy and degeneracy of the spin model with exact diagonalization. Financial support from CPNq is acknowledged by C.S.deF. (project No. 435665/2016-2), E.M. (307041/2017-4) and R.G.P. (303298/2019-7). E.M. and R.G.P also acknowledge Capes/Cofecub 0899/2018. V.S.deC. thanks the financial support from FAPESP and CAPES under Grants Nos. 2016/05069-7 and 88887.469170/2019-00, respectively.  Research at IIP-UFRN is supported by Brazilian ministries MEC and MCTIC. 
\end{acknowledgments}

\appendix

\section{Majorana fermion representation \label{app:Majorana}}
In this appendix, we discuss the transformation of the Majorana fermions in the spin liquid under time reversal.  

At each site, we can combine the six Majorana fermions $(\eta^\alpha,\theta^\alpha)$ to define three complex fermions:\bea
c^x&=&\frac12(\eta^x-i\theta^x),\nonumber\\
c^y&=&\frac12(\eta^y-i\theta^y),\nonumber\\
c^z&=&\frac12(\eta^z-i\theta^z),
\eea
which obey $\{c^\alpha,(c^\beta)^\dagger\}=\delta^{\alpha\beta}$. 
States in this Fock   space are specified by  $|n_x,n_y,n_z\rangle$, where $n_\alpha\in\{0,1\}$ are the fermion occupation numbers.  Thus, this representation generates 8 states, which is twice the size of the physical Hilbert space for spin $3/2$. We can represent the four eigenstates of $J^z$ by identifying\bea
&\left|\frac32\right\rangle&=|0,0,0\rangle\equiv |\varnothing\rangle,\nonumber\\
&\left|\frac12\right\rangle&=|1,1,0\rangle=(c^x)^\dagger(c^y)^\dagger|\varnothing\rangle,\nonumber\\
&\left|-\frac12\right\rangle&=|0,1,1\rangle=(c^y)^\dagger(c^z)^\dagger|\varnothing\rangle,\nonumber\\
& \left|-\frac32\right\rangle&=|1,0,1\rangle=(c^x)^\dagger(c^z)^\dagger|\varnothing\rangle.
\eea
This corresponds to imposing the parity constraint\be
(2n_x-1)(2n_y-1)(2n_z-1)=-1,
\ee
which in terms of Majorana fermions becomes Eq. (\ref{constraint}).

Since time reversal acts on the spin-$3/2$ states as\bea
&T\left|\frac32\right\rangle=\left|-\frac32\right\rangle,\qquad T\left|\frac12\right\rangle=-\left|-\frac12\right\rangle,\nonumber\\
&T\left|-\frac12\right\rangle=\left|\frac12\right\rangle,\qquad T\left|-\frac32\right\rangle=-\left|\frac32\right\rangle,
\eea
we   postulate that time reversal is equivalent to a particle-hole transformation for fermions $c^x$ and $c^z$, such that  $Tc^{x,z}T^{-1}=(c^{x,z})^\dagger$ and $T|0,0,0\rangle=|1,0,1\rangle$.  We can then check that \bea
&T\left|-\frac32\right\rangle&=T(c^x)^\dagger (c^z)^\dagger |\varnothing\rangle\nonumber\\
&&=c^xc^zT|\varnothing\rangle\nonumber\\
&&=c^xc^z(c^x)^\dagger(c^z)^\dagger|\varnothing\rangle\nonumber\\
&&=-|\varnothing\rangle,
\eea
as expected. Likewise, it is straightforward to verify       the time reversal transformation of the $\left|\pm\frac12\right\rangle$ states. The rule of particle-hole transformation for $c^x$ and $c^z$ but not for $c^y$ is equivalent to applying  complex conjugation and taking $\theta^y\mapsto -\theta^y$, as mentioned in Sec. \ref{sec:THs}.

\section{Ground state degeneracy of the time-reversal-symmetric quadrupolar spin liquid \label{App:degeneracy}}
In this appendix, we discuss the ground state degeneracy of the Hamiltonian  in Eq. (\ref{Hspin}).   

First, consider the case $\lambda=0$, in which  the Hamiltonian commutes with the operators $\tau_j^y$ on every site. Using the representation of pseudospin and pseudo-orbital operators, we can write the Hamiltonian in the form \be
H_s(\lambda=0)=16K\sum_{\gamma=x,y,z}\sum_{\langle jl\rangle_\gamma } \tau^y_j \tau^y_l s_j^\gamma s_l^\gamma.
\ee
We can replace the conserved quantities by their eigenvalues, $\tau^y_j=\frac12 \xi_j$, where $\xi_j\in\{\pm1\}$, and obtain\be
H_s(\lambda=0)=4\sum_{\gamma=x,y,z}\sum_{\langle jl\rangle_\gamma }K_{jl}s_j^\gamma s_l^\gamma,\label{KjlKitaev}
\ee
where $K_{jl}=K \xi_j \xi_l $.  The Hilbert space can then  be divided into sectors of $\{\xi_j\}$. Note that the product $\xi_j \xi_l $ determines the sign of the exchange coupling $K_{jl}$ between the pseudospins  on the bond $\langle jl\rangle_\gamma$. For instance, if we fix $\xi_j=1$ $\forall j$, the model reduces to the homogeneous spin-$1/2$ Kitaev model. Moreover, the Kitaev model with opposite sign for the coupling is obtained in the sector with $\xi_j=1$  for all sites $j$ in sublattice A and $\xi_l=-1$ for all sites $l$ in sublattice B.  Importantly,  $\xi_j$ and $\xi_l $ are   eigenvalues of  local physical operators, as opposed to gauge variables that arise in parton constructions. 

We now show that every eigenstate in a given sector  is degenerate with a state in another sector where the set $\{\xi_j\}$ differs only by a local change of the eigenvalues in a single bond. This implies  an extensive degeneracy that scales with the number of bonds, which is manifested in the Majorana fermion representation through the presence of zero-energy flat bands in the spectrum of Fig. \ref{Bulk_Dispersions_SL}.

Consider that we start  in a given sector, where   Hamiltonian (\ref{KjlKitaev}) has a set of eigenstates $\{|\Psi_n(\{\xi_j\})\rangle\}$,   and change  to a different sector by inverting the sign of $\xi_j$ and $\xi_l$ on a single $z$ bond. This transformation does not affect the sign of $K_{jl}$ on that bond, but it flips the sign of the couplings of sites $j$ and $l$ to their nearest neighbors on $x$ and $y$ bonds. However, this sign change can be removed by applying a local $\pi$ rotation around the $z$ axis, which takes $s^{x,y}_j\mapsto -s^{x,y}_j$ and $s^{x,y}_l\mapsto -s^{x,y}_l$. This means that the Hamiltonian in the new sector is related to the original one by a canonical transformation, and to any eigenstate in the original sector there corresponds an eigenstate in the new sector with the same energy.  

For $\lambda\neq0$, the $\tau_j^y$ operators are no longer conserved, but one can verify that the local two-site operators $\tau_j^z\tau^z_ls_j^zs^z_l$, with $j,l$ nearest neighbors on $z$ bonds, commute with the Hamiltonian. In the Majorana fermion representation, these operators are written as $\frac{i}4u_{\langle jl\rangle_z}\theta_j^z\theta_l^z$. Thus, their conservation law is associated  with the fact that the $\theta_j^z$ Majorana fermion does not appear in the Hamiltonian in Eq. (\ref{H_spins_majorana}). Let us denote the eigenvalues of $\tau_j^z\tau^z_ls_j^zs^z_l$ by $\frac1{16}\xi_{jl}$, where $\xi_{jl}\in\{\pm1\}$. In a sector with fixed $\{\xi_{jl}\}$, we can write\be
s_j^zs^z_l=\xi_{jl}\tau_j^z\tau^z_l.
\ee
The Hamiltonian in Eq. (\ref{Hspin}) can then be expressed in the form
\bea
H_s&=&16K\sum_{\gamma=x,y}\sum_{\langle jl\rangle_\gamma } \tau^y_j \tau^y_l s_j^\gamma s_l^\gamma\nonumber\\
&&+16K\sum_{\langle jl\rangle_z } \xi_{jl}\tau^y_j \tau^y_l \tau_j^z \tau_l^z-2\lambda\sum_j\tau^z_j\nonumber\\
&=&16K\sum_{\gamma=x,y}\sum_{\langle jl\rangle_\gamma } \tau^y_j \tau^y_l s_j^\gamma s_l^\gamma\nonumber\\
&&-4K\sum_{\langle jl\rangle_z } \xi_{jl}\tau^x_j \tau^x_l -2\lambda\sum_j\tau^z_j. \label{xijl}
\eea
The argument for the degeneracy now is similar to the one given above. Suppose we flip the sign of a single $\xi_{jl}$. We can   map the  Hamiltonian in the new sector to the original one by applying a unitary transformation that takes $\tau_j^{x,y}\mapsto-\tau_j^{x,y}$ and $s_j^{x,y}\mapsto-s_j^{x,y}$ on site $j$ only. Note that the $\pi $ rotation acting on both pseudospins is necessary to preserve the sign of the first term in Eq. (\ref{xijl}). We conclude that also for $\lambda\neq0$ there is an extensive  degeneracy in the spectrum of $H_s$.

\section{Effective action  \label{App1}}

In this appendix, we provide details about the calculation of the effective action discussed  in  Sec.  \ref{sec:Instabilities-induced-by}. The term $\mathcal{H}_c(\mathbf{k})$ in Eq. (\ref{action}) is the Hamiltonian matrix obtained by Fourier transforming $H_c$ in Eq. \eqref{eq:Hconduction}, which written as matrix in sublattice space becomes\begin{widetext}
\begin{equation}
\mathcal{H}_c(\mathbf{k})=\frac{1}{2}
\begin{pmatrix} 
\Delta_0(\mathbf{k})(\sigma^x\rho^z+\sigma^z\rho^x) & if(\mb k)\left[
t(\sigma^x\rho^z+\sigma^z\rho^x)+w(\sigma^z\rho^z-\sigma^x\rho^x)\right] \\
-if^*(\mathbf{k})\left[
t(\sigma^x\rho^z+\sigma^z\rho^x)+w(\sigma^z\rho^z-\sigma^x\rho^x)
\right] & -\Delta_0(\mathbf{k})(\sigma^x\rho^z+\sigma^z\rho^x)\end{pmatrix}.
\end{equation}
\end{widetext}
In this form, it is easy to see that $\mathcal{H}_c(\mathbf{k})$ already contains  triplet pairing correlations. However,   we want to address the  contributions  contained in the electron  self-energy, which  has the form \begin{align}
\Sigma_{\text{F/AF}}(\mathbf{k},i\omega_n)&=\mathcal{V}_{\text{F/AF}}\frac{1}{i\omega_n-\mathcal{H}_s(\mathbf{k})}\mathcal{V}^\dagger_{\text{F/AF}},\label{Eq_mom_freq_SE}
\end{align}
where $\mc H_s(\mb k)$ is given in Eq. (\ref{Hsk}) and $\mc V_{\text{F/AF}}$ are hybridization matrices between the  Balian-Werthamer spinor and the Majorana fermions in the spin liquid. 

\subsection{Uniform configuration}

For the F state, the hybridization matrix   is
\begin{equation}
\mathcal{V}^\dagger_\text{F}=\frac{iJ_K}{\sqrt2}
\begin{pmatrix} 
0 & 0 & 0 & 0 & 0 & 0 & 0 & 0 \\
0 & 0 & 0 & 0 & 0 & 0 & 0 & 0 \\
0 & -1 & 1 & 0 & 0 & 0 & 0 & 0 \\
0 & 0 & 0 & 0 & 0 & -1 & 1 & 0 
\end{pmatrix}.
\end{equation}
Substituting the above expression into Eq. \eqref{Eq_mom_freq_SE}, we obtain the  self-energy 
\bea
	\Sigma_\text{F}(\mathbf{k},i\omega_n)&=&\frac{J^2_K}{2}(\mathbb{1}_4-\boldsymbol\sigma\cdot\boldsymbol\rho)\otimes\mc M(\mb k,i\omega_n),
\eea
where we    introduce the matrix in sublattice space\be
\mc M(\mb k,i\omega_n)=\begin{pmatrix} 
	i\Omega(\mathbf{k},i\omega_n) & i\Gamma(\mathbf{k},i\omega_n) \\
	-i\Gamma^*(\mathbf{k},i\omega_n) & i\Omega(\mathbf{k},i\omega_n)
	\end{pmatrix},
\ee
with the following functions:
\begin{align}
\Omega(\mathbf{k},i\omega_n)&=-\frac{\omega_n[\omega^2_n+|g(\mathbf{k})|^2+\lambda^2]}{2[(\omega^2_n+\lambda^2)^2+|g(\mathbf{k})|^2\omega^2_n]},\\
\Gamma(\mathbf{k},i\omega_n)&=\frac{\lambda^2g(\mathbf{k})}{2[(\omega^2_n+\lambda^2)^2+|g(\mathbf{k})|^2\omega^2_n]}.
\end{align}
As done in Ref. \cite{Coleman1994}, we decompose $\Sigma_\text{F}(\mathbf{k},i\omega_n)$ into three terms:
\begin{equation}
\Sigma_\text{F}(\mathbf{k},i\omega_n)=\Sigma^\text{N}_\text{F}(\mathbf{k},i\omega_n)+\Sigma^\text{SC}_\text{F}(\mathbf{k},i\omega_n)+\Sigma^\text{RE}_\text{F}(\mathbf{k},i\omega_n),
\end{equation}
where
\bea
\Sigma^\text{N}_\text{F}(\mathbf{k},i\omega_n)&=& \frac{J^2_K}{2}\mathbb{1}_4\otimes \mc M(\mb k,i\omega_n),\nonumber\\
\Sigma^\text{SC}_\text{F}(\mathbf{k},i\omega_n)&=&-\frac{J^2_K}{4}(\sigma^+ \rho^{-}+\sigma^- \rho^{+})\otimes\mc M(\mb k,i\omega_n),\nonumber\\
\Sigma^\text{RE}_\text{F}(\mathbf{k},i\omega_n)&=&-\frac{J^2_K}{2}\sigma^z\rho^z\otimes\mc M(\mb k,i\omega_n).
\eea
Expressing the action  in terms of the two-component spinors $\psi_{A/B}(\mathbf{k},i\omega_n)$, we find the superconducting action and pairing function written  in Eqs. (\ref{FK_SC_action}) and (\ref{FK_pairing}), respectively.

Proceeding rather similarly, the induced RE action in the F state is obtained as
\bea
\delta\mathcal{S}_\text{RE}&=& \sum_{\mathbf{k},\omega_n}\sum_{b,b'}\big[\psi^\dagger_b(\mathbf{k},i\omega_n)\chi_{bb'}^\text{RE}(\mathbf{k},i\omega_n)\psi_{b'}(\mathbf{k},i\omega_n)\nonumber\\
&&+\text{H.c.}\big],
\eea
where
\begin{align}
\chi_{\text{AA}}^\text{RE}(\mathbf{k},i\omega_n)&=-\frac{iJ^2_K}{4}\frac{\omega_n(\omega^2_n+|g(\mathbf{k})|^2+\lambda^2)\sigma^z}{(\omega^2_n+\lambda^2)^2+|g(\mathbf{k})|^2\omega^2_n},\nonumber\\
\chi_{\text{AB}}^\text{RE}(\mathbf{k},i\omega_n)&=\frac{iJ^2_K}4\frac{\lambda^2g(\mathbf{k})\sigma^z}{(\omega^2_n+\lambda^2)^2+|g(\mathbf{k})|^2\omega^2_n},\nonumber\\
\chi_{\text{BA}}^\text{RE}(\mathbf{k},i\omega_n)&=-\frac{iJ^2_K}4\frac{\lambda^2g^*(\mathbf{k})\sigma^z}{(\omega^2_n+\lambda^2)^2+|g(\mathbf{k})|^2\omega^2_n},\nonumber\\
\chi_{\text{BB}}^\text{RE}(\mathbf{k},i\omega_n)&=-\frac{iJ^2_K}4\frac{\omega_n(\omega^2_n+|g(\mathbf{k})|^2+\lambda^2)\sigma^z}{(\omega^2_n+\lambda^2)^2+|g(\mathbf{k})|^2\omega^2_n}.
\end{align}
The above order parameters indicate the emergence of a nonlocal  order, which is completely polarized along the $z$ direction.

\subsection{Staggered configuration}

The procedure to calculate the induced SC and RE actions for the AF state  follows the same steps described in the last subsection. The hybridization matrix in this case is \begin{equation}
\mathcal{V}^\dagger_\text{AF}=\frac{iJ_K}{\sqrt2} \begin{pmatrix} 
0 & 0 & 0 & 0 & 0 & 0 & 0 & 0 \\
0 & 0 & 0 & 0 & 0 & 0 & 0 & 0 \\
0 & -1 & 1 & 0 & 0 & 0 & 0 & 0 \\
0 & 0 & 0 & 0 & 1 & 0 & 0 & 1 
\end{pmatrix}.
\end{equation}
 The self-energy in Eq. \eqref{Eq_mom_freq_SE} has the form 
\begin{equation}
\Sigma_\text{AF}(\mathbf{k},i\omega_n)=\Sigma^\text{N}_\text{AF}(\mathbf{k},i\omega_n)+\Sigma^\text{SC}_\text{AF}(\mathbf{k},i\omega_n)+\Sigma^\text{RE}_\text{AF}(\mathbf{k},i\omega_n),
\end{equation}
where\begin{widetext}
\begin{align}
\Sigma^\text{N}_\text{AF}(\mathbf{k},i\omega_n)&=\frac{J^2_K}{2}\mathbb{1}_4\otimes i\Omega(\mathbf{k},i\omega_n) \nonumber\\
\Sigma^\text{SC}_\text{AF}(\mathbf{k},i\omega_n)&=\frac{J^2_K}{2}\begin{pmatrix} -i\Omega(\mathbf{k},i\omega_n)(\rho^x\sigma^x+\rho^y\sigma^y)& 
\Gamma(\mathbf{k},i\omega_n) (\rho^y+i\rho^x\sigma^z)\\ 
\Gamma^*(\mathbf{k},i\omega_n)(\rho^y-i\rho^x\sigma^z)& 
i\Omega(\mathbf{k},i\omega_n) (\rho^x\sigma^x+\rho^y\sigma^y)\end{pmatrix},\nonumber\\
\Sigma^\text{RE}_\text{AF}(\mathbf{k},i\omega_n)&=\frac{J^2_K}{2}\begin{pmatrix} -i\Omega(\mathbf{k},i\omega_n) \rho^z\sigma^z  & 
-\Gamma(\mathbf{k},i\omega_n) (\sigma^y+i\rho^z\sigma^x)\\ 
-\Gamma^*(\mathbf{k},i\omega_n) (\sigma^y-i\rho^z\sigma^x) & 
i\Omega(\mathbf{k},i\omega_n) \rho^z\sigma^z\end{pmatrix}.
\end{align}
\end{widetext}
These   results allow us to write down the induced SC action and the  pairing functions shown in Eqs. (\ref{AFK_SC_action}) and (\ref{AFK_pairing}).

Lastly, the   RE action for the AF state  is given by
\bea
\delta\tilde{\mathcal{S}}_\text{RE}&=& \sum_{\mathbf{k},\omega_n}\sum_{b,b'}\big[\psi^\dagger_b(\mathbf{k},i\omega_n)\tilde{\chi}_{bb'}^\text{RE}(\mathbf{k},i\omega_n)\psi_{b'}(\mathbf{k},i\omega_n)\nonumber\\
&&+\text{H.c.}\big],
\eea
where
\begin{align}
\tilde{\chi}_{\text{AA}}^\text{RE}(\mathbf{k},i\omega_n)&=-\frac{iJ^2_K}4\frac{\omega_n(\omega^2_n+|g(\mathbf{k})|^2+\lambda^2)\sigma^z}{(\omega^2_n+\lambda^2)^2+|g(\mathbf{k})|^2\omega^2_n},\nonumber\\
\tilde{\chi}_{\text{AB}}^\text{RE}(\mathbf{k},i\omega_n)&=\frac{iJ^2_K}2\frac{\lambda^2g(\mathbf{k})\sigma^-}{(\omega^2_n+\lambda^2)^2+|g(\mathbf{k})|^2\omega^2_n},\nonumber\\
\tilde{\chi}_{\text{BA}}^\text{RE}(\mathbf{k},i\omega_n)&=-\frac{iJ^2_K}2\frac{\lambda^2g^*(\mathbf{k})\sigma^+}{(\omega^2_n+\lambda^2)^2+|g(\mathbf{k})|^2\omega^2_n},\nonumber\\
\tilde{\chi}_{\text{BB}}^\text{RE}(\mathbf{k},i\omega_n)&=\frac{iJ^2_K}4\frac{\omega_n(\omega^2_n+|g(\mathbf{k})|^2+\lambda^2)\sigma^z}{(\omega^2_n+\lambda^2)^2+|g(\mathbf{k})|^2\omega^2_n}.
\end{align}
As a result,   the above order parameters contain  polarization vectors which are no longer restricted to the $z$ direction.

\bibliographystyle{apsrev4-1_control}
\bibliography{./Ref}

\end{document}